\documentclass[trackchanges,resetfootnote]{aastex701}
\usepackage[normalem]{ulem}

\begin{document}

\title{Photometric Characterisation and Evolutionary State Investigation of the Low Mass-Ratio Contact Binary FASTT 1558}

\author[orcid=0009-0005-4311-0810]{Anurag Baruah}
\affiliation{Department of Physics, Indian Institute of Technology Hyderabad, Kandi, Sangareddy 502284, India}
\email[show]{ph24resch11002@iith.ac.in}  

\author[orcid=0000-0002-5900-9785]{Mayukh Pahari} 
\affiliation{Department of Physics, Indian Institute of Technology Hyderabad, Kandi, Sangareddy 502284, India}
\email{mayukh@phy.iith.ac.in}

\author[orcid=0009-0005-3844-3426]{Kumar Pranshu}
\affiliation{Aryabhatta Research Institute of Observational sciencES (ARIES), Manora Peak, Nainital 263001, India}
\email{pranshu@aries.res.in}

\author[orcid=0009-0000-8543-7292]{Nihal Mohamed Thalakkott}
\affiliation{Department of Physics, Indian Institute of Technology Hyderabad, Kandi, Sangareddy 502284, India}
\email{ph24mscst11021@iith.ac.in}

\author[orcid=0000-0003-1637-267X]{Kuntal Misra}
\affiliation{Aryabhatta Research Institute of Observational sciencES (ARIES), Manora Peak, Nainital 263001, India}
\email{kuntal@aries.res.in}

\author[orcid=0000-0000-0000-0000]{Ikyatha Tummalapalli}
\affiliation{Department of Astronomy, Osmania University, Hyderabad 500007, India}
\email{ikyatha.tummalapalli@gmail.com}

\begin{abstract}
Contact binaries with low mass ratios, as potential progenitors of stellar mergers, offer important insights into the physical processes governing mass and energy transfer. In this work, we present the first spectro-photometry based evolutionary investigation of FASTT 1558, a totally eclipsing, low mass-ratio contact binary. We combine \textsc{TESS} and \textsc{ASAS-SN} photometry with phase-resolved spectroscopy obtained with the 3.6m Devasthal Optical Telescope. The lightcurve modeling with \texttt{PHOEBE} shows that the system has an exceptionally low secondary-to-primary mass ratio of $q=0.1544\pm0.0013$, with a deep contact degree of $f=0.61\pm0.01$. Our analysis yields a total system mass of $1.92\pm0.31\,M_\odot$, while the distance to the system is estimated to be $648.45\pm29.85$ pc. Analysis of long-term eclipse timings reveals an apparent orbital period decrease of $\dot{P}=-(3.03\pm0.26) \times 10^{-7}$ ${\rm day\,yr^{-1}}$, which may be associated with mass-transfer from the primary to the secondary on a thermal timescale. The spectra show phase-dependent H$\alpha$ and Ca\textsc{ii} equivalent widths, indicating a dynamically evolving chromosphere. Binary evolution calculations with \texttt{MESA} terminates through L2 overflow before reaching the observed low mass-ratio. Despite its extreme mass ratio, the system is presently Darwin stable, with $J_{\rm spin}/J_{\rm orb}=0.15\pm0.02$. FASTT 1558 therefore provides a robust observational benchmark for evolutionary models of low mass-ratio contact binaries.
\end{abstract}

\section{INTRODUCTION} 
Binary or multiple stellar systems constitute the majority of the stellar population in the universe \citep{1983ARA&A..21..343A, 2010ApJS..190....1R, 2013ARA&A..51..269D, 2017ApJS..230...15M, 2018maeb.book.....P, 2021MNRAS.507.5013M}. In the Milky Way galaxy, nearly 30-40\% of low-mass stars are found in binary configurations \citep{1991A&A...248..485D, 2008MNRAS.389..869E, 2010ApJS..190....1R, 2013ARA&A..51..269D, 2021MNRAS.507.5013M}. Owing to the precise determination of their fundamental physical parameters, close binary systems serve as important laboratories for testing theories of stellar structure and evolution. Various statistical studies have shown that contact binaries of W Ursae Majoris type are the most common among close binary systems with relatively slow evolutionary behavior \citep{1948HarMo...7..249S, 1998AJ....116.2998R, 2002PASP..114.1124R, 2001AcA....51..259S, 2006AcA....56..199S}.

In these systems, material and energy are redistributed between the stars through the inner Lagrangian point (L1), allowing energy to be transported from the hotter component to the cooler one via a common envelope \citep{1968ApJ...151.1123L, 1968ApJ...153..877L}. The heating of the less massive secondary can drive Thermal Relaxation Oscillations \citep[TROs;][]{1976ApJ...205..208L, 1976ApJ...205..217F, 1977MNRAS.179..359R, 2001MNRAS.328..914Q, 2003MNRAS.342.1260Q, 2005ApJ...629.1055Y}, causing the direction of mass transfer to alternate and the system to repeatedly transition between contact and near-contact configurations. These oscillations drive the binary system toward progressively lower mass ratios \citep{2001AJ....122.1007R, 2006AcA....56..199S, Stepien2012, 2023A&A...672A.176P}. As a result, the system may eventually encounter the Darwin instability \citep{darwin1879viii, 1980A&A....92..167H}, potentially leading to the merger of the binary components \citep{2011A&A...528A.114T, 2014ApJ...788...22P, 2015AJ....150...69Y, 2026arXiv260828066L}. Due to their extremely short orbital periods and correspondingly high binding energies, the common envelope cannot be entirely expelled during the final stages of evolution, making coalescence of the components highly probable \citep{2000ARA&A..38..113T, 2001A&A...369..939W, 2026ApJ...998..232J}. 

Theoretically, the critical mass ratio for Darwin instability has been predicted to lie in the range \(q_{\rm inst}\sim0.09\)–\(0.038\), depending on stellar structure, rotation, metallicity, and the adopted gyration radii \citep{1995ApJ...444L..41R, 2006MNRAS.369.2001L, 2009MNRAS.394..501A, 2021MNRAS.501..229W, 2024NatSR..1413011Z}. However, observations have revealed contact binaries with mass ratios substantially below these theoretical limits, reaching \(q\simeq0.024\) in TYC 3801-1529-1 \citep{2026ApJ...998..108P}, highlighting the unresolved discrepancy between theoretical stability limits and the observed population of extremely low mass-ratio contact binaries. Although several theoretical models have been proposed to explain the evolutionary nature of contact binaries \citep{1976ApJ...205..208L, 1976ApJ...205..217F, 1977ApJ...215..851W, 2005ApJ...629.1055Y, 2006AcA....56..199S, Stepien2012, 2013MNRAS.430.2029Y, 2020MNRAS.492.4112Z, 2026ApJS..282...17Z}, the formation mechanisms and long-term evolution of these systems are still subjects of ongoing debate \citep{1968ApJ...151.1123L, 2016ApJ...817..133Z, 2020MNRAS.492.4112Z, 2023ApJ...958...84K, 2023A&A...672A.175F, 2024RAA....24d5018P, 2024MNRAS.535.2494W}. Further investigations of both known and poorly studied systems are therefore crucial for constraining their evolutionary pathways and improving our understanding of their formation and ultimate fate.

In this study, we present the first optical spectro-photometric analysis of the low mass-ratio, low-mass contact binary FASTT 1558, discovered by \citet{1998AJ....115..296H} using the Flagstaff Astrometric Scanning Transit Telescope (FASTT), together with an investigation of its evolutionary status through stellar evolution modeling. FASTT 1558 has appeared repeatedly in many statistical studies such as Catalina Sky Survey \citep[CSS;][]{2014ApJS..213....9D}, All Sky Automated Survey \citep[ASAS;][]{1997AcA....47..467P}, Northern Sky Variability Survey \citep[NSVS;][]{2004AJ....127.2436W,2006AJ....131..621G}, The Two Micron All-Sky Survey \citep[2MASS;][]{2003yCat.2246....0C} and more recently in Transiting Exoplanet Survey Satellite \citep[TESS;][]{2015JATIS...1a4003R,2022ApJS..258...16P}, All Sky Automated Survey for SuperNovae \citep[ASAS-SN;][]{2014AAS...22323603S, 2018MNRAS.477.3145J} and Large Sky Area Multi-Object Fiber Spectroscopic Telescope \citep[LAMOST;][]{2015RAA....15.1095L, 2020RAA....20..163Q} survey. However, no detailed spectro-photometric study along with long-term evolution of the system has been reported to date. Some basic information on FASTT 1558 is presented in Table \ref{tab:info}. 
\begin{table}[h]
    \centering
    \caption{Basic information on FASTT 1558.}
    \begin{tabular}{lr}
    \hline
    Parameter & Value\\
    \hline
    \hline  
    R.A. & 22$^h$33$^m$55$^s$.84\\
    Dec. & +00$^d$24$^m$50$^s$.4\\
    P$_\mathrm{orb}$ (days) & 0.477305\\
    V (mag) & 12.21\\
    Spectral Type & F0\\
    \hline
    \end{tabular}
    \begin{flushleft}
        Note. The spectral type is taken from \citet{2020RAA....20..163Q} and all the other parameter values are from \citet{2014ApJS..213....9D}.
    \end{flushleft}
    \label{tab:info}
\end{table}

The spectroscopic observation and analysis are presented in Section \ref{sec:spec}. In Section \ref{sec:lc} we present the photometric observation and analysis. Section \ref{sec:evo} discusses evolutionary state of the binary through numerical modeling. Finally, the implications of our findings and the conclusions of this work are presented in Section \ref{sec:diss-con}.
\section{Spectroscopic OBSERVATION AND DATA Analysis}
\label{sec:spec}
\subsection{Spectroscopic Reduction}
Spectroscopic observations are conducted using the ARIES-Devasthal Faint Object Spectrograph and Camera \citep[ADFOSC;][]{2019arXiv190205857O} mounted on the 3.6-meter Devasthal Optical Telescope \citep[DOT;][]{2018BSRSL..87...29K}. The instrument is equipped with a 4K $\times$ 4K deep-depletion CCD, providing a plate scale of 0.2" per pixel with 2 $\times$ 2 on-chip binning \citep{2023JATIS...9a8002P}.

A total of 36 spectra are acquired, each with 300-second exposures at different orbital phases on 16 November 2025. The 676R grism is employed in combination with a 1.5" slit, covering a wavelength range of approximately 450-880 nm. The spectroscopic data are reduced using standard procedures within the Image Reduction and Analysis Facility \citep[\texttt{IRAF 2.18.1};][]{1986SPIE..627..733T} package, including bias subtraction, flat-field correction, spectral extraction, and wavelength calibration. Wavelength calibration is performed using Fe-Ne arc lamp exposures, while flux calibration is achieved through observations of the spectrophotometric standard white dwarf GD 50. The resulting spectral resolution is approximately R $\simeq$ 700 at 700 nm, calculated using the Fe-Ne arc lamp. The fully calibrated spectra are subsequently continuum-normalised using the \texttt{specutils v2.3.1} package of \texttt{Astropy 7.2} \citep{2013A&A...558A..33A}, enabling consistent comparison of line profiles and chromospheric features across orbital phases. 
\subsection{Spectroscopic Analysis}
\label{sec:sp}
\begin{figure}[h]
    \centering
    \includegraphics[width=\linewidth]{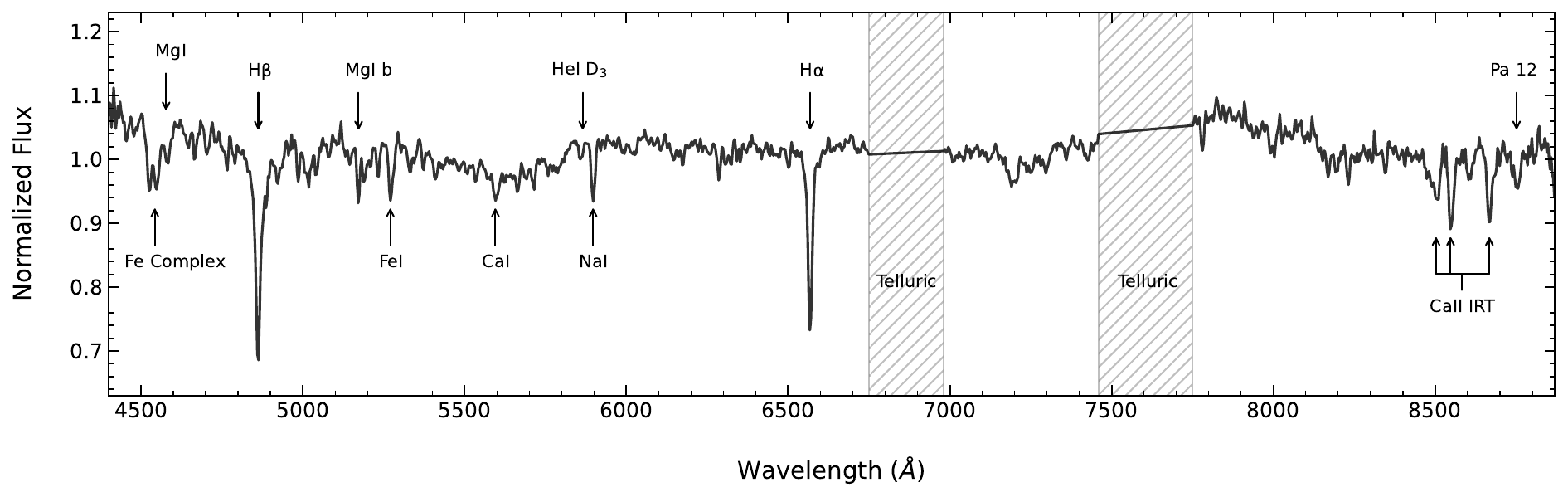}
    \caption{Optical Spectra of FASTT 1558 obtained with ADFOSC spectrograph in 3.6 m DOT, ARIES, highlighting prominent absorption lines and telluric bands.}
    \label{fig:fastt_spectra}
\end{figure}

To avoid contamination from telluric absorption features (molecular oxygen, O$_2$), we mask the wavelength regions 6750-6980~\AA~and 7460-7750~\AA. Figure~\ref{fig:fastt_spectra} illustrates the identified absorption features and masked telluric bands in the continuum normalized FASTT 1558 spectrum. 
\subsubsection{Stellar Parameter Determination}
To determine the fundamental atmospheric parameters of FASTT 1558, we employ the University of Lyon Spectroscopic analysis Software \citep[\texttt{ULySS};][]{2009A&A...501.1269K} package. It performs pixel-by-pixel $\chi^2$ minimisation between an observed spectrum and a model spectrum constructed from interpolated stellar libraries. The \texttt{elodie32\_flux\_tgm.fits} interpolator within \texttt{ULySS} is employed, which is constructed from the ELODIE 3.2 spectral library, to generate model spectra in the wavelength range 4500-6500~\AA, corresponding to the specified atmospheric parameters: effective temperature ($T_{\mathrm{eff}}$), surface gravity ($\log g$), and metallicity ([Fe/H]) \citep{2011A&A...525A..71W}. In this wavelength range, regions affected by strong chromospheric activity (H$\beta$ and Na\textsc{i} core) are excluded from the fitting process to prevent bias in the parameter determination. The weighted mean values of the atmospheric parameters along with the weighted scatters obtained from all the 36 spectra are given in Table \ref{tab:atm_param}. The uncertainties of the weighted means do not account for systematic effects associated with the spectral resolution, the adopted template library, continuum normalization, or the composite nature of the spectrum of the binary system. These effects may contribute additional uncertainties to the derived atmospheric parameters and should be considered when interpreting the results. Due to the low resolution of the spectra, no significant shift in wavelength is detected. The parameters reported by the LAMOST and Gaia DR3 surveys are also presented in the Table \ref{tab:atm_param} for comparison.  

\begin{table}[h]
    \centering
    \caption{Atmospheric parameters of FASTT 1558. The mean values from all the 36 spectra are presented along with the values reported by various other surveys.}
    \begin{tabular}{lcccccc}
    \hline
    Observatory & Observations Date & $T_\mathrm{eff}$ & $\log \mathrm{g}$ & [Fe/H] & RV & Reference\\
     &   (UTC) & (K) & [cm/s$^2$] & [dex] & (km/s) &\\
    \hline
    \hline
    DOT/ADFOSC & 2025-11-16 & 6713.52 $\pm$ 34.52 & 4.089 $\pm$ 0.048 & -0.005 $\pm$ 0.037 & -- & This Study\\
    LAMOST & 2017-12-06 & 6612.49 $\pm$ 14.50 & 4.079 $\pm$ 0.024 & -0.125 $\pm$ 0.013 & -2.42 $\pm$ 13.50 & \citet{2020RAA....20..163Q}\\
    LAMOST & 2015-10-01 & 6607.54 $\pm$ 10.17 & 4.089 $\pm$ 0.015 & -0.111 $\pm$ 0.008 & 23.24 $\pm$ 11.30 & \citet{2020RAA....20..163Q}\\
    Gaia DR3 & -- & 6266.21$^{+24.53}_{-18.40}$ & 3.94 $\pm$ 0.01 & -0.384 $\pm$ 0.03 & 14.73 $\pm$ 15.35 & 
    \citet{Gaia2021}\\
    \hline
    \end{tabular}
    \label{tab:atm_param}
\end{table}
\subsubsection{Chromospheric Activity and Detection of Inverse P-Cygni Profile}
\begin{figure}[h]
    \centering
    \includegraphics[width=0.5\linewidth]{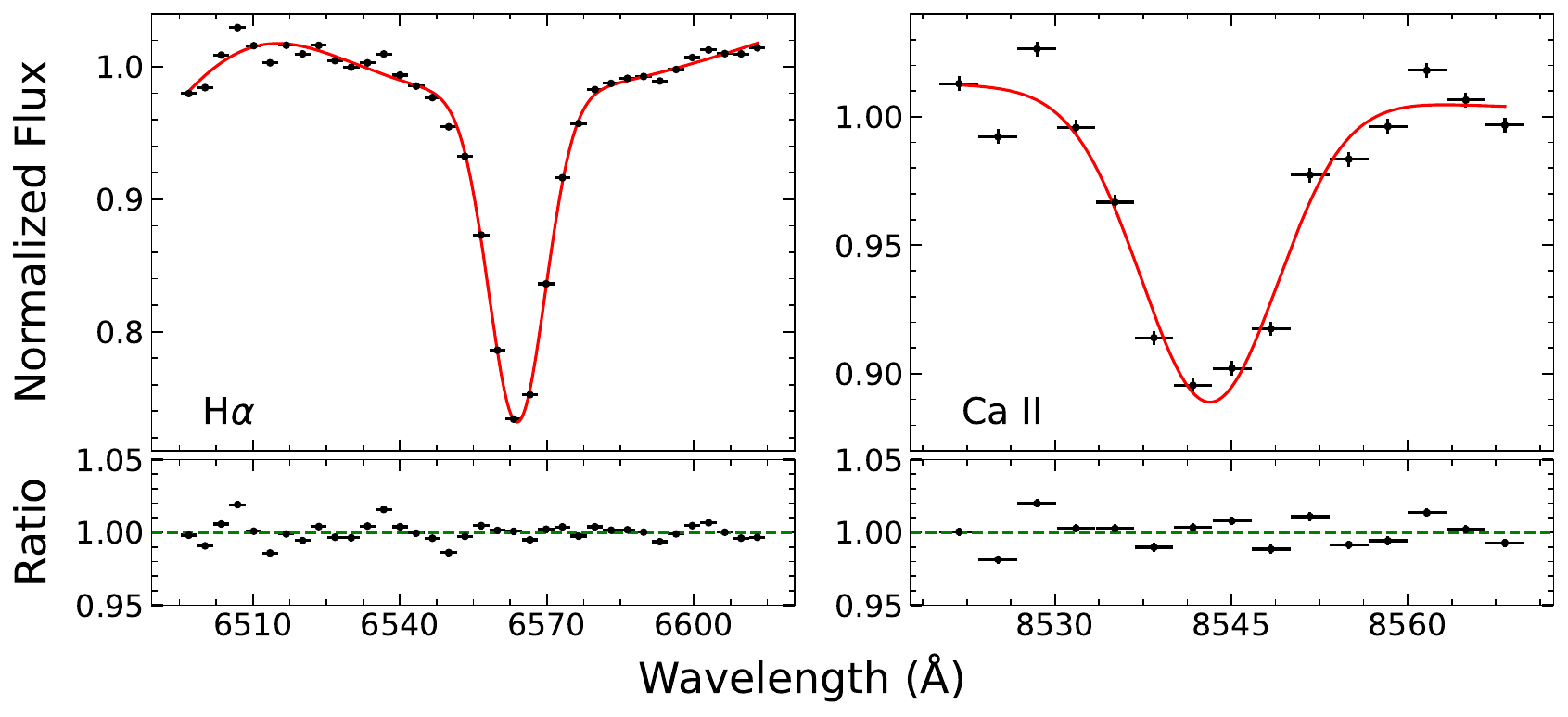}
    \\
    \vspace{0.3cm}
    \includegraphics[width=0.5\linewidth]{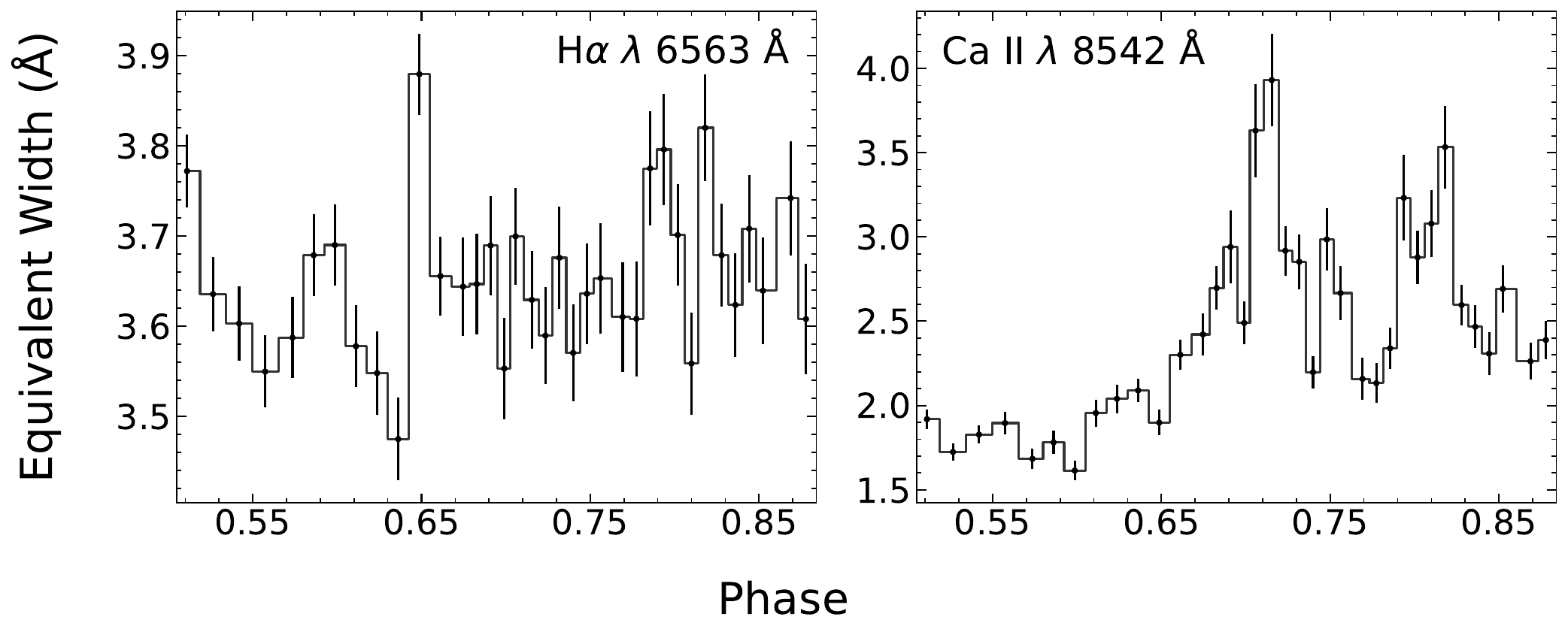}
    \caption{\textit{Top}: H$\alpha$ (left) and Ca\textsc{ii} (right) absorption feature of FASTT 1558 at phase 0.52. In each panel, the upper section shows the observed data (black points) along with the best-fit model (red line), while the lower section displays the data-to-model ratio. \textit{Bottom}: Evolution of equivalent widths (EW) for H$\alpha$ (left) and Ca\textsc{ii} (right) lines with phase in FASTT 1558.}
    \label{fig:figure-ew}
\end{figure}

For line diagnostics, we focus on the H$\alpha$ line at 6563~\AA~and the strongest component of the Ca\textsc{ii} Infrared Triplet (IRT) at 8542~\AA~as these are known chromospheric lines in F type stars. The continuum-normalised spectra around the H$\alpha$ and Ca\textsc{ii} lines are fitted using a Gaussian profile to model the absorption component. To account for any broad hump-like residuals, a combination of linear and Lorentzian components is adopted for the H$\alpha$ region, while only a linear component is used for the Ca\textsc{ii} line. The resulting fitted profiles along with the ratios of fitting at phase 0.52 are shown in top panels of Figure \ref{fig:figure-ew}. From the fitted models, we derive the corresponding equivalent widths (EWs) and their variations with orbital phase is shown in bottom panels of Figure \ref{fig:figure-ew}. All the fitted spectra for H$\alpha$ and Ca\textsc{ii} lines are presented in Figure \ref{fig:h-a} and \ref{fig:ca-ii} respectively, along with their EW values in Table \ref{tab:p-ew}.

\begin{figure}[h]
    \centering
    \includegraphics[width=0.5\linewidth]{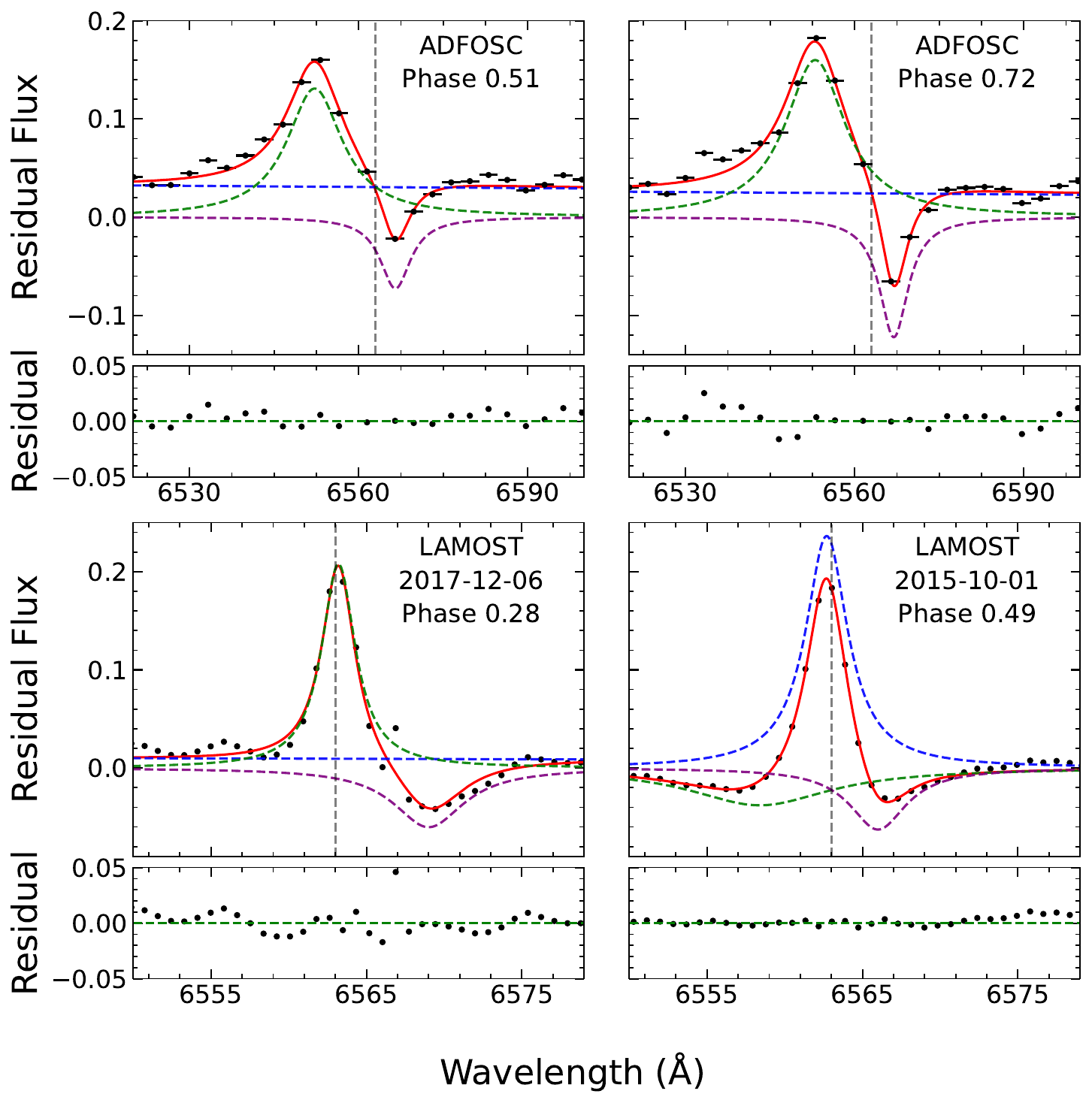}
    \caption{H$\alpha$ residual spectra of FASTT 1558 from ADFOSC (top panels) and LAMOST (bottom panels). The black points represent the observed spectra, while the solid red curves show the best-fit composite models consisting of Lorentzian and linear components. The vertical grey dashed line marks the rest wavelength of H$\alpha$. The lower sub-panels display the residuals of the fits.}
    \label{fig:h_alpha_chro}
\end{figure}

To isolate the chromospheric contribution from the underlying photosphere, we employ the spectral subtraction technique implemented in \texttt{iSTARMOD} \citep{2026AJ....171...15L}. The high-resolution PHOENIX model spectrum \citep{2013A&A...553A...6H}, generated using the mean parameter values obtained from the \texttt{ULySS} fitting (see Table \ref{tab:atm_param}), and convolved to match the resolution of our ADFOSC spectra, is adopted as the template. Continuum normalisation is performed consistently for both observed and template spectra prior to subtraction to avoid artificial residual structures. The subtraction procedure is carried out over the full spectral range, excluding strong telluric bands to prevent contamination. The residual spectra of H$\alpha$ for all the 36 observations, shows an inverse P-Cygni profile, which is also visible in the two available LAMOST observations. We have shown this feature with a fitted model (combination of linear and Lorentzian functions) in Figure \ref{fig:h_alpha_chro} along with residuals. The redshifted component of the ADFOSC spectra at phase 0.51 is centered at $6570.53 \pm 0.12$ \AA, causing a shift of $7.73 \pm 0.12$ \AA\ from the H$\alpha$ line center at 6562.8 \AA. Similar shifts are observed in all the spectra (see Table \ref{tab:p-ew}). We have used \texttt{scipy.optimize.curve\_fit} \citep{2020NaMet..17..261V} for modeling the feature and the quoted uncertainties are derived from the diagonal elements of the covariance matrix returned by the fitting procedure. These uncertainties however, do not account for the contributions from wavelength calibration uncertainties or possible instrumental systematic effects.

\section{Photometric Observation and data analysis}
\label{sec:lc}
\subsection{Photometric Reduction}
Photometric data from TESS and ASAS-SN are used in this study. TESS, launched in April 2018, is a space-based photometric survey mission optimized for exoplanet detection. It employs four wide-field cameras providing a combined field of view of approximately 24$^{\circ} \times$ 96$^{\circ}$ and observes through a broad-band filter spanning roughly 600-1000 nm. FASTT 1558 (TIC 293106548) is observed by TESS in Sectors 42, 70, 82 and 92. The light curves are downloaded from the Mikulski Archive for Space Telescopes data archive (MAST\footnote{\url{https://archive.stsci.edu}}) using the Python package \texttt{lightkurve} \citep{2018ascl.soft12013L}. The light curves processed through the Science Processing Operations Center (SPOC) pipeline are available with exposure times of 600 seconds in Sectors 42 and 82, and 200 seconds in Sector 70. Both Simple Aperture Photometry (SAP) and Pre-search Data Conditioning SAP (PDCSAP) fluxes are provided in TESS data products. Sector 92 was excluded from our study as only Quick Look Pipeline (QLP) light curves are available at the time of the analysis, whereas systematics-corrected PDCSAP fluxes from the SPOC pipeline are used for the other sectors.

On the other hand, ASAS-SN is a ground-based time-domain photometric survey that provides long-term photometric monitoring in the V and g bands with a typical cadence of a few days. FASTT 1558 (ASASSN-V J223355.87+002450.3) has been monitored by ASAS-SN from 2013 to 2019. In the present work, we use the V-band photometric data available at the ASAS-SN Variable Stars Database\footnote{\url{https://asas-sn.osu.edu/variables}}.
\subsection{Photometric Analysis}
\subsubsection{Light Curve Solutions}
\begin{figure}[h]
    \centering
    \includegraphics[width=0.5\linewidth]{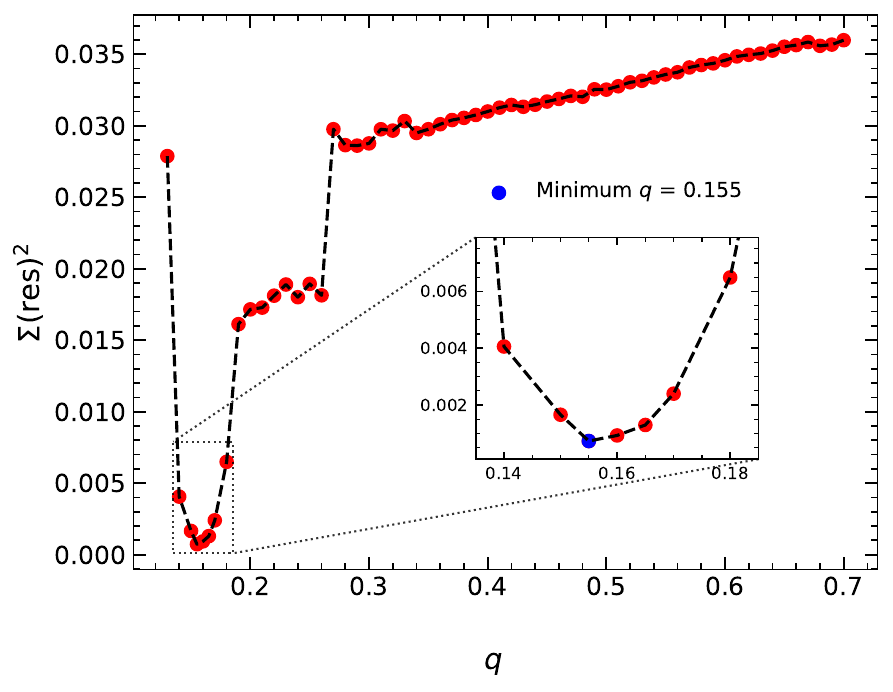}
    \caption{\textit{q-search}: sum of the squared residuals as a function of the mass ratio.}
    \label{fig:q-search}
\end{figure}

We have used the Python code PHysics Of Eclipsing BinariEs \citep[\texttt{PHOEBE 2.4.};][]{2005ApJ...628..426P, 2020ApJS..250...34C} for light-curve analysis. Due to the absence of multiepoch radial-velocity (RV) data, \textit{q-search} technique is used for the estimation of mass-ratio from the photometric data. As FASTT 1558 is a totally eclipsing contact binary, its light-curve morphology can provide reliable estimate of the mass ratio \citep{2001AJ....122.1007R, 2003CoSka..33...38P, 2005Ap&SS.296..221T}. The effective temperature of the star eclipsed at the primary minimum is fixed at T$_1$ = T$_\mathrm{eff}$ = 6713 K (Table \ref{tab:atm_param}), as derived from our ADFOSC spectral analysis. We fix the bolometric albedos A$_1$ = A$_2$ = 0.5 \citep{1969AcA....19..245R} and gravity-darkening coefficients g$_1$ = g$_2$ = 0.32 \citep{1967ZA.....65...89L}, assuming convective envelopes (as T $<$ 7200 K). The limb-darkening coefficients are interpolated for logarithmic law from van Hamme's table \citep{1993AJ....106.2096V}. Then, we vary the mass-ratio (q parameter) from 0.13 to 0.7 in steps of 0.01 and run the \texttt{PHOEBE} program corresponding to each value. In this process, other parameters like secondary star temperature (T$_2$), primary component surface potential ($\Omega_1$), primary component luminosity (L$_1$) and inclination angle (i) are set as free parameters. The sum of squared residuals ($\Sigma (res)^2$) obtained corresponding to each fits are plotted against q in Figure \ref{fig:q-search}, which shows that the solution is converged at q = 0.155 corresponding to minimum $\Sigma (res)^2$ (a step size of 0.005 is used around the minima). The corresponding parameters obtained in \textit{q-search} are our initial estimates.

\begin{figure}[h]
    \centering
    \includegraphics[width=0.92\linewidth]{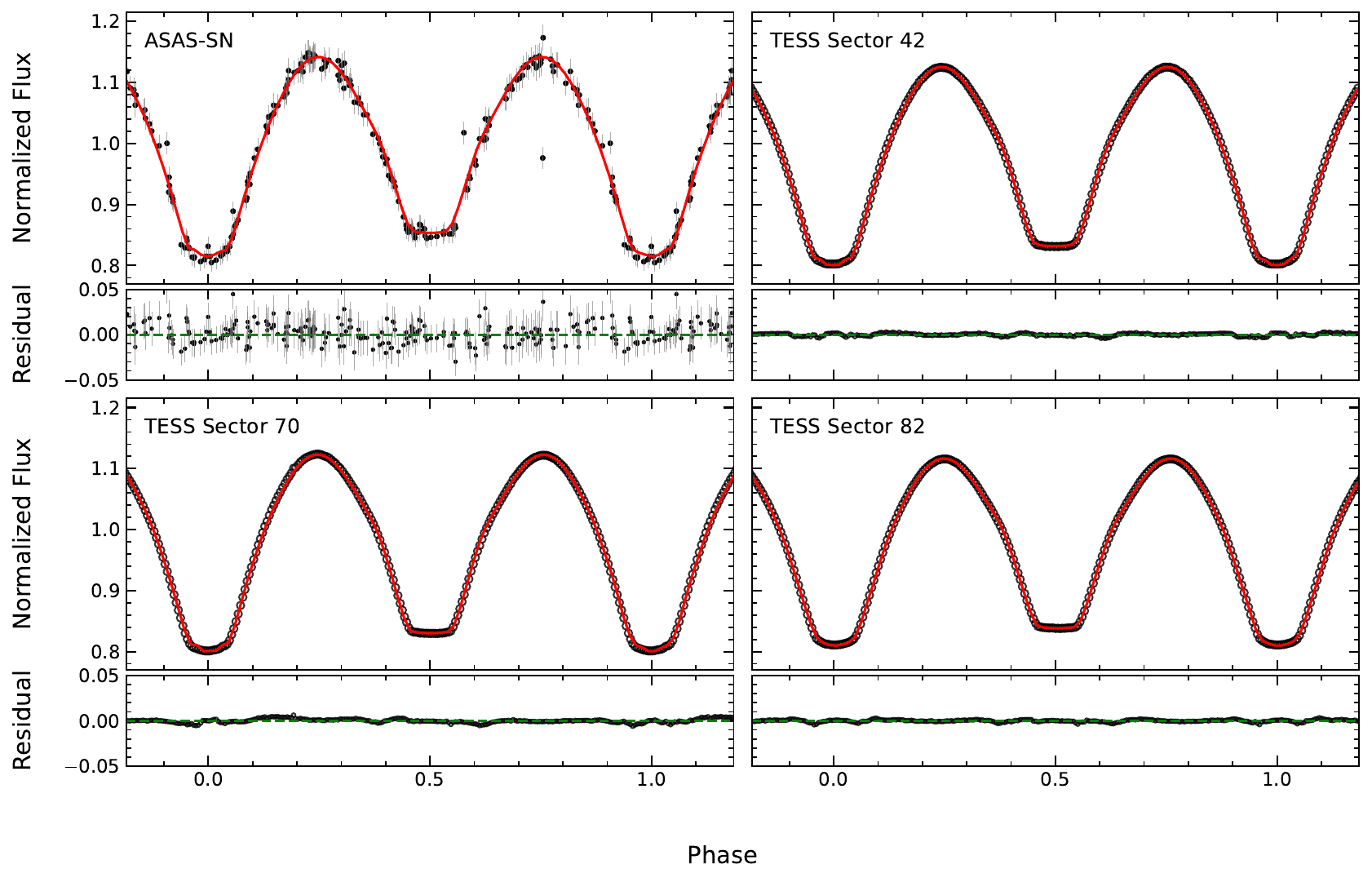}
    \caption{Light-curve modelling of FASTT 1558. \textit{Top of each panel}: observed (black circles) and theoretical (red solid lines) light curves calculated from \texttt{PHOEBE} model. \textit{Bottom of each panel}: residuals between observed and theoretical light curves.}
    \label{fig:lc}
\end{figure}
\begin{figure}[h]
    \centering
    \includegraphics[width=0.49\linewidth]{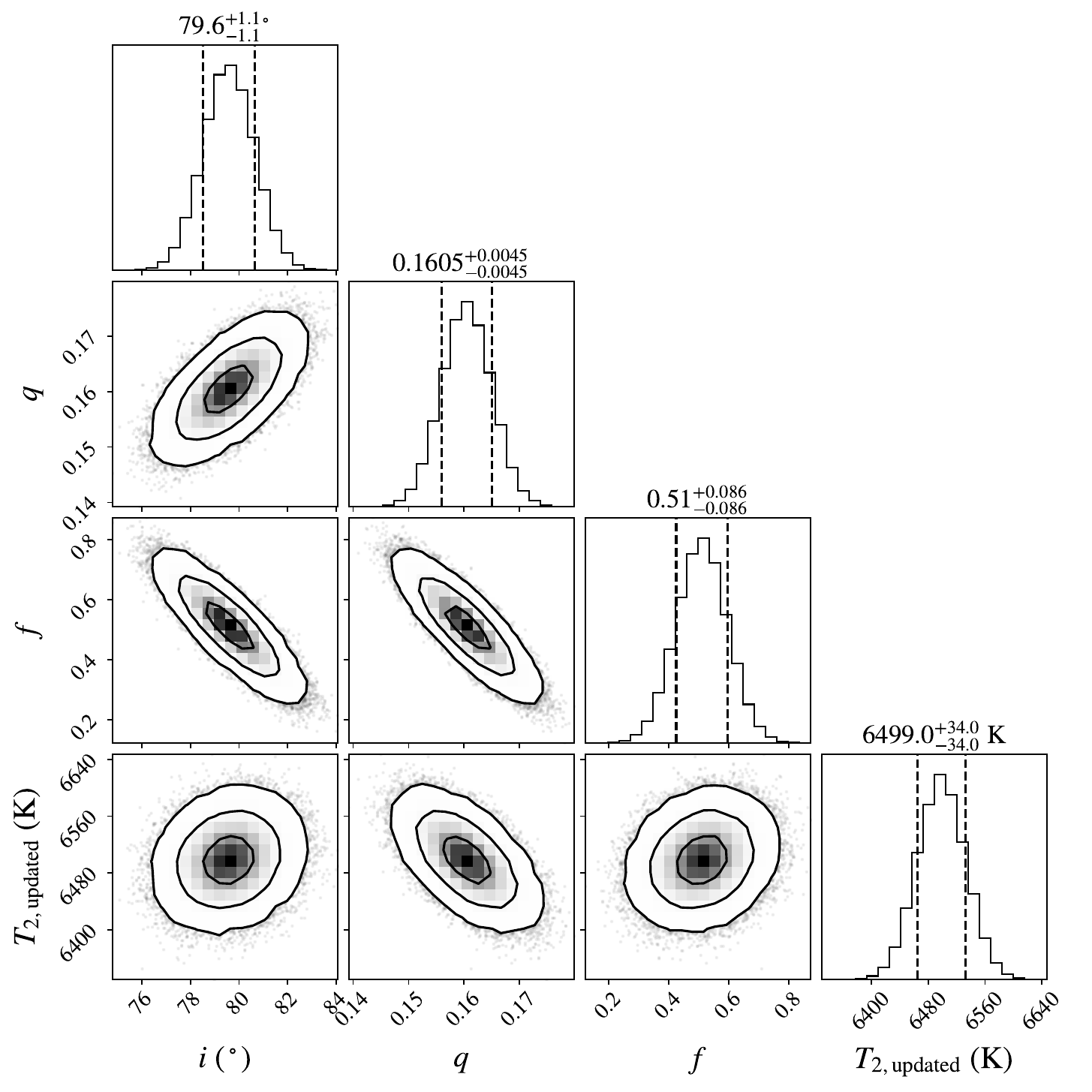}
    \hfill
    \includegraphics[width=0.49\linewidth]{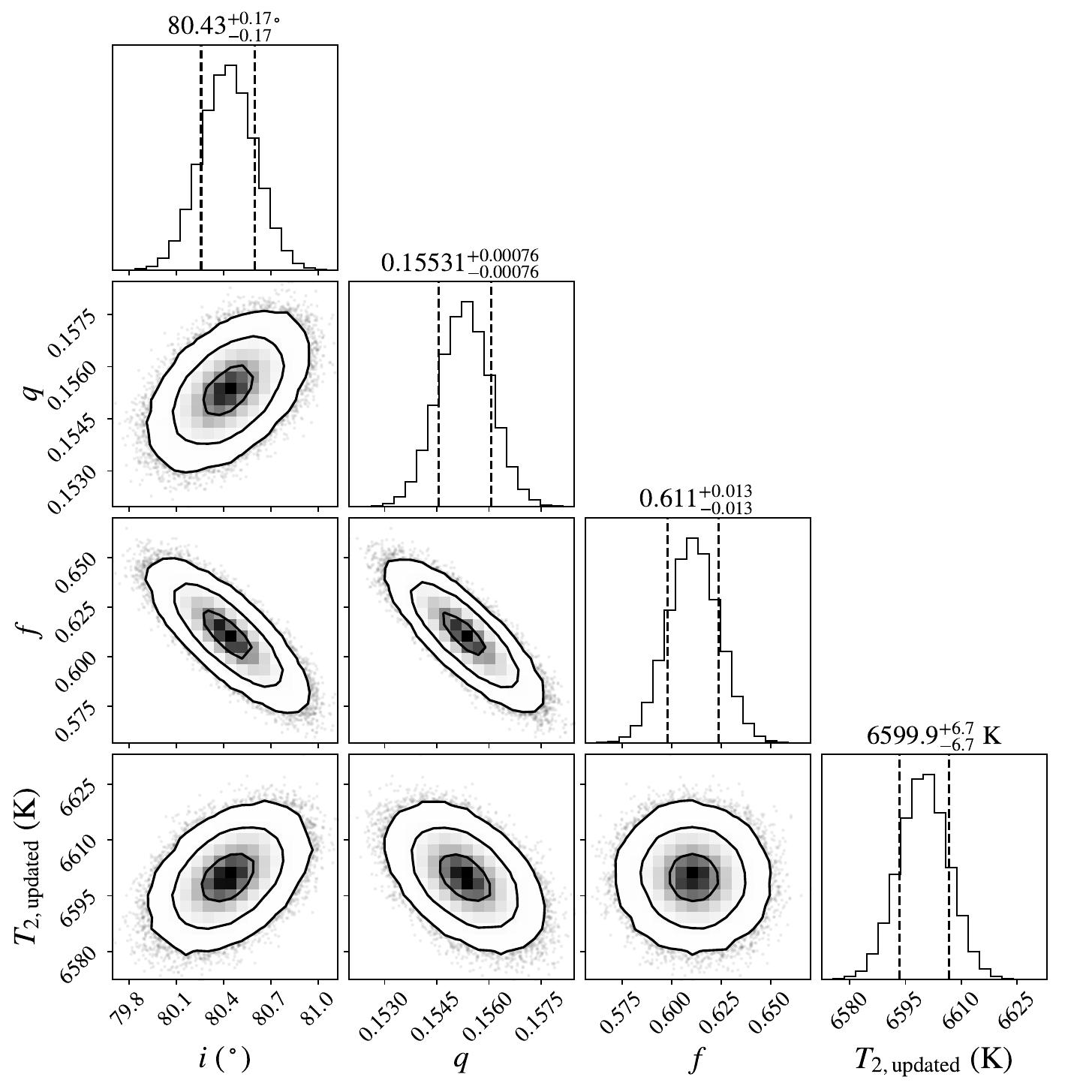}
    \caption{Posterior distributions of the modelled parameters derived from MCMC analysis of the ASAS-SN (left) and TESS sector-70 (right) light curves.}
    \label{fig:corner}
\end{figure}

Since the initially adopted value of T$_1$, derived from spectroscopic analysis, represents a surface-weighted effective temperature of both components, we recalculate the individual component temperatures using the following relations \citep{2003A&A...404..333Z, 2013AJ....146..157C},

\begin{equation}
    T_1,\rm{updated} = \left( \frac{(1+k^2)T_{\mathrm{eff}}^4}{1+k^2\alpha^4} \right)^{0.25}, \ \ T_2,\rm{updated} = T_1,\rm{updated} \times \alpha
\end{equation}
where T$_{\mathrm{eff}}$ denotes the effective temperature from spectroscopic analysis, while $k = \frac{r_2}{r_1}$ and $\alpha = \frac{T_2}{T_1}$ are obtained from the initial \texttt{PHOEBE} estimates. Using updated values of temperatures, $T_1,\rm{updated}$ and $T_2,\rm{updated}$ (see Table \ref{tab:lc}), we rerun the \texttt{PHOEBE} program. The final parameter values and the associated uncertainties of the four datasets are estimated independently using a Markov Chain Monte Carlo \citep[MCMC;][]{2013ascl.soft03002F} analysis approach, employing 32 walkers and 2000 steps. After excluding the initial 90 iterations as burn-in, we examine the acceptance fractions and integrated autocorrelation times for all four datasets. The diagnostics show similar behaviour across the datasets, with mean acceptance fractions of approximately 0.54 and the ratio of chain length to autocorrelation time of about 50. Figure \ref{fig:lc} shows the observed (black dots) and modeled (red lines) phased light curves and the posterior probability distributions of the parameters with their multivariate Gaussian representation for ASAS-SN and TESS 70 LCs are presented in Figure \ref{fig:corner}. The determined photometric parameters are listed in Table \ref{tab:lc}.
\begin{table}[h]
    \caption{Light curve solutions for FASTT 1558.}
    \begin{tabular}{lccccc}
    \hline
    Parameter & ASAS-SN & TESS-42 & TESS-70 & TESS-82 & Average\\
    \hline
    \hline
    $T_1,\rm{updated}$(K)$^{*}$ & 6748 & 6748& 6748& 6748& 6748\\
    $T_2,\rm{updated}$(K) & 6499 $\pm$ 34 & 6628.4 $\pm$ 3.8 & 6599.9 $\pm$ 6.7 & 6692 $\pm$ 13 & 6624.63 $\pm$ 23.68\\
    $i$ ($^\mathrm{o}$) & 79.60 $\pm$ 1.10 & 80.21 $\pm$ 0.14 & 80.43 $\pm$ 0.17 & 79.36 $\pm$ 0.27 & 80.16 $\pm$ 0.34\\
    $q = M_2/M_1$ & 0.1605 $\pm$ 0.0045 & 0.1542 $\pm$ 0.0006 & 0.1553 $\pm$ 0.0007 & 0.1508 $\pm$ 0.0015 & 0.1544 $\pm$ 0.0013\\
    $f$ & 0.51 $\pm$ 0.09 & 0.61 $\pm$ 0.01 & 0.61 $\pm$ 0.01 & 0.57 $\pm$ 0.03 & 0.61 $\pm$ 0.01\\
    $l_1/l_{tot}$ & 0.837 $\pm$ 0.017 & 0.835 $\pm$ 0.021 & 0.840 $\pm$ 0.016 & 0.837 $\pm$ 0.019 & 0.837 $\pm$ 0.009\\
    $l_2/l_{tot}$ & 0.163 $\pm$ 0.020 & 0.165 $\pm$ 0.021 & 0.160 $\pm$ 0.016 & 0.163 $\pm$ 0.017 & 0.163 $\pm$ 0.009\\
    $r_{1 \ (mean)}$ & 0.567 $\pm$ 0.004 & 0.562 $\pm$ 0.002 & 0.573 $\pm$ 0.005 & 0.567 $\pm$ 0.002 & 0.567 $\pm$ 0.001\\
    $r_{2 \ (mean)}$ & 0.257 $\pm$ 0.004 & 0.260 $\pm$ 0.001 & 0.263 $\pm$ 0.005 & 0.257 $\pm$ 0.002 & 0.259 $\pm$ 0.001\\
    \hline
    \end{tabular}
    \begin{flushleft}
        $^*$: fixed parameter during the modelling. $f$ denotes the fillout factor and $r_{1,2}$ are the fractional radii.
    \end{flushleft}
    \label{tab:lc}
\end{table}

For determination of the average values of the parameters, we use weighted mean and scatter method. The weighted means are calculated as defined in the following equations, where \(x_i\) denotes the individual measurements with uncertainties \(\sigma_i\).
\begin{equation}
    \bar{x} = \frac{\sum_{i=1}^{N} w_i x_i}{\sum_{i=1}^{N} w_i} \ , \  w_i = \frac{1}{\sigma_i^2}
\end{equation}
The scatter among the individual measurements around the weighted mean is quantified using the weighted variance, defined as
\begin{equation}
    s_w^2 = \frac{\sum_{i=1}^{N} w_i (x_i-\bar{x})^2}{\sum_{i=1}^{N} w_i}
\end{equation}
The weighted scatter \(s_w\), characterizes the dispersion of the individual measurements around the weighted mean, with measurements having smaller uncertainties contributing more strongly to the calculation.
\subsubsection{Absolute Physical Parameters Estimation}
Due to the lack of radial velocity curves, the absolute physical parameters of the binary system can be derived using either the Gaia-distance-based \citep[GD;][]{2019RAA....19...14K} method or empirical relations \citep[ER;][]{2020Ap&SS.365...71L}. Using statistical analysis, \citet{2021AJ....162...13L} showed that the parameters derived using the GD method can be unreliable, particularly for distant contact binaries \citep{2026AJ....171..254Y}. Therefore, we used the empirical relation of \citet{2024PASP..136b4201P},
\begin{equation}
    a = (0.372^{+0.113}_{-0.114}) + (5.914^{+0.272}_{-0.298}) \times P
\end{equation}
which links the orbital period (P) and semi-major axis ($a$) for contact binaries, based on a homogeneous sample of 414 systems with periods shorter than 0.7 days.

\begin{figure}[h]
    \centering
    \includegraphics[width=\linewidth]{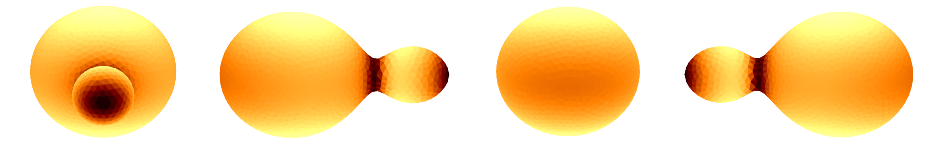}
    \caption{3D representations of the components of FASTT 1558 at phases 0.00, 0.25, 0.50 and 0.75 respectively.}
    \label{fig:3D}
\end{figure}

The radii of the stars $R_{1,2}$ are calculated using the mean fractional radii presented in Table \ref{tab:lc},
\begin{equation}
    R_{1,2} = r_{1,2} \times a
\end{equation}
The total mass is be determined by applying Kepler's third law,
\begin{equation}
    M = a^3/P^2
\end{equation}
where a is in astronomical units and period P is in years. Using the values of radii and temperatures from Table \ref{tab:lc}, the luminosities of the two components are calculated,
\begin{equation}
    \frac{L_{1,2}}{L_{\odot}} = \left( \frac{R_{1,2}}{R_\odot}\right)^2 \left(\frac{T_{1,2}}{T_\odot}\right)^4 
\end{equation}
Additionally, the orbital angular momentum is evaluated based on the formulation presented by \citet{2006MNRAS.373.1483E},
\begin{equation}
    J_{orb} = 1.24 \times 10^{52} \times M^{5/3} \times P^{1/3} \times q \times (1 + q)^{-2}
\end{equation}
All the absolute parameters with their associated uncertainties are presented in Table \ref{tab:params} and Figure \ref{fig:3D} displays the 3D representation of the components at different orbital phases.
\begin{table}[h]
    \centering
    \caption{The physical parameters of FASTT 1558.}
    \begin{tabular}{lc}
    \hline
    Parameters & Values\\
    \hline
    \hline
    $M_1/\mathrm{M_{\odot}}$ & 1.66 $\pm$ 0.27\\
    $M_2/\mathrm{M_{\odot}}$ & 0.26 $\pm$ 0.04\\
    $R_1/\mathrm{R_{\odot}}$ & 1.81 $\pm$ 0.10\\
    $R_2/\mathrm{R_{\odot}}$ & 0.83 $\pm$ 0.04\\
    $L_1/\mathrm{L_{\odot}}$ & 6.13 $\pm$ 0.66\\
    $L_2/\mathrm{L_{\odot}}$ & 1.19 $\pm$ 0.13\\
    $a/\mathrm{R_{\odot}}$ & 3.19 $\pm$ 0.17\\
    $\log \mathrm{J_{orb}}$ (cgs)& 51.52 $\pm$ 0.12\\
    \hline 
    \end{tabular}
    \label{tab:params}
\end{table}

We also derive the relative energy transfer rates between the components as suggested by \citet{1981ApJ...245..650M},
\begin{equation}
    U_1 = \frac{r^2 t^4 - q^\alpha}{1 + q^\alpha}
    \label{eq-ET-rate}
\end{equation}

\begin{equation}
    U_2 = \log \left[ \frac{r^2 t^4 - q^\alpha}{q^\alpha (1 + r^2 t^4)} \right]
\end{equation}
where $r = R_2 /R_1$, $t = T_2 /T_1$ correspond to radius and temperature ratios, and $\alpha$ = 3.42 represents the exponent of the mass luminosity relation. The nuclear luminosity of each star is estimated using the single-star empirical relation provided by \citet{1991Ap&SS.181..313D},
\begin{equation}
    L_0 \simeq 1.03 \, M^{3.42}, \quad 0.1 \le M \le 120
\end{equation}
These values are presented in Table \ref{tab:energy-rate} and are consistent with the theoretical fits for $U_1$, $U_2$ and $\log(L_{10}+L_{20})$ presented by \citet{2026PASP..138c4203S}.

\begin{table}[h]
    \centering
    \caption{Nuclear luminosities and relative energy transfer rates of the components of FASTT 1558.}
    \begin{tabular}{cccc}
    \hline
    $L_{0,1}$ $(L_\odot)$ & $L_{0,2}$ $(L_\odot)$ & $U_{1}$ & $U_{2}$ \\
    \hline
    \hline
    5.87 $\pm$ 3.21 & 0.01 $\pm$ 0.005 & 0.19 $\pm$ 0.02 & 1.99 $\pm$ 0.06\\
    \hline
    \end{tabular}
    \label{tab:energy-rate}
\end{table}
\subsubsection{Determination of the distance to FASTT 1558}
The distance to the system is estimated using the luminosities derived from the light curve (LC) solution along with bolometric corrections (BC). This method provides a physically consistent estimate based on stellar parameters. The bolometric magnitude of each component is calculated using:
\begin{equation}
M_{\mathrm{bol},i} = M_{\mathrm{bol},\odot} - 2.5 \log \left( \frac{L_i}{L_\odot} \right)
\end{equation}
where $M_{\mathrm{bol},\odot} = 4.74$ is the solar bolometric magnitude, and $L_i$ is the luminosity of each component. The bolometric correction in the V-band is obtained using the polynomial calibration \citep{1996ApJ...469..355F, 2010AJ....140.1158T}:
\begin{equation}
BC_V = a + b x + c x^2 + d x^3 + e x^4
\end{equation}
where $x = \log T_{\mathrm{eff}}$, and $a, b, c, d, e$ are coefficients corresponding to the appropriate temperature range. The absolute visual magnitude for each component is therefore:
\begin{equation}
M_{V,i} = M_{\mathrm{bol},i} - BC_i
\end{equation}
We use the total absolute magnitude of the binary system in the distance modulus relation,
\begin{equation}
m_V - M_V = 5 \log d - 5 + A_V,
\end{equation}
where $m_V = 11.86$ is the apparent magnitude at maximum brightness and $A_V = 3.1 \times E(B-V) = 0.242$ (color excess, $E(B-V)$ of 0.078) is the interstellar extinction in the V-band. These values are adopted from ASAS-SN data. Therefore, the distance is calculated as $d$ = 648.45 $\pm$ 29.85 pc, which is consistent with Gaia dsphot measurements 609 $\pm$ 6 pc \citep{Anders2022}. In the uncertainty of the derived distance, the uncertainties associated with interstellar extinction have not been accounted for and this value may therefore underestimate the total uncertainty.
\subsubsection{Orbital Period Variations}
\label{sec:o-c}
\begin{figure}[h]
    \centering
    \includegraphics[width=0.5\linewidth]{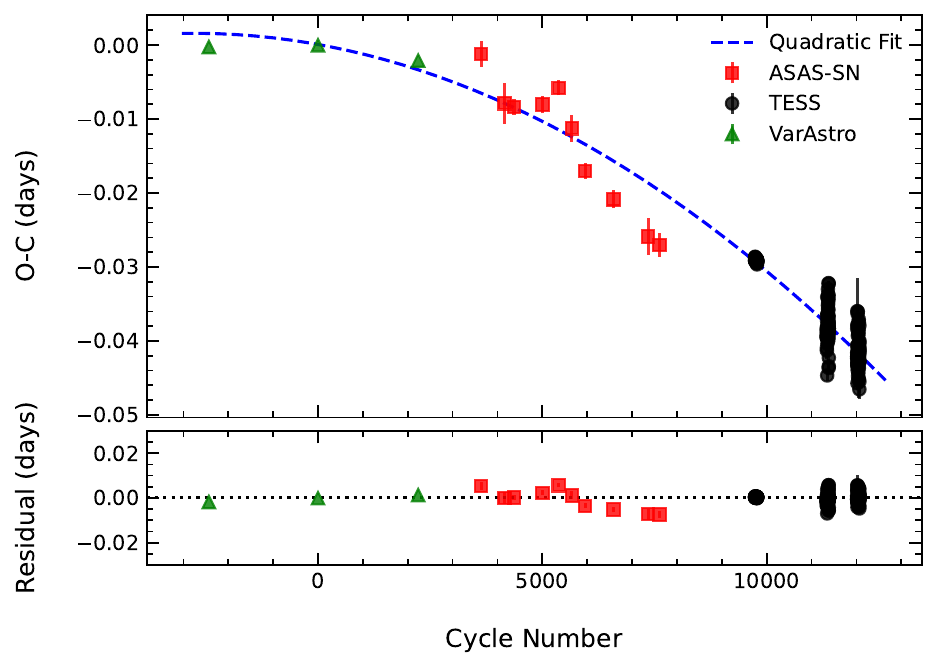}
    \caption{O-C diagram illustrating the period variation of the system.}
    \label{fig:o-c}
\end{figure}

The temporal variation in the orbital period of contact binaries provides useful information about mass transfer rate, presence of third body and other characteristics. To quantify this variation, we determine  times of minima using the TESS and ASAS-SN light curves. Four additional minima are adopted from VarAstro\footnote{\url{https://var.astro.cz/en/Stars}}. Since TESS used BJD time system, the minima in HJD are converted to BJD using \texttt{Astropy 7.2}. The TESS minima are determined using the KvW method \citep{1956BAN....12..327K}, incorporating the improved error estimation described by \citet{2020Galax...9....1D}. As the ASAS-SN observations spanning 2013-2019 are not continuous, the data are divided into 150 day intervals as prescribed by \citet{2026arXiv260717650A}. The observations in each interval are matched with a template (phased TESS 82 light curve) to get the minima. The following linear equation is used to determine the calculated values of times of minima,
\begin{equation}
    BJD = BJD_0 + P \times E
\end{equation}
where E being the cycle/epoch number. The values of epoch ($BJD_0$) and period are taken from VarAstro (see Table \ref{tab:o-c1} and \ref{tab:o-c2}). The Observed minus Calculated (O-C) diagram is shown in Figure \ref{fig:o-c}, which has a downward parabolic trend. Therefore, we fit the curve with a quadratic polynomial and the fitting results are given in Table \ref{tab:o-c}. From the fitted coefficients, the derivative of the orbital period for FASTT 1558 is calculated. However, the apparent long-term decrease in the orbital period, (3.03 $\pm$ 0.26) $\times$ 10$^{-7}$ day yr$^{-1}$, may represent only a segment of a long-term cyclic variations, potentially caused by a third body \citep{1952ApJ...116..211I, 2005ASPC..335..103P} or magnetic activity cycles \citep{1992ApJ...385..621A}. Further observational data are necessary to verify this interpretation. The calculated O-C values for FASTT 1558 are presented in Table \ref{tab:o-c1} and \ref{tab:o-c2}.

\begin{table}[h]
    \caption{Fitted and calculated parameters from the O-C diagram.}
    \centering
    \begin{tabular}{cc}
    \hline
    Parameters  & Values\\
    \hline
    \hline
    $a_0$ & (0.0000874 $\pm$ 0.00106)\\
    $a_1$ & (-1.09 $\pm$ 0.26) $\times$ 10$^{-6}$\\
    $a_2$ & (-1.98 $\pm$ 0.17) $\times$ 10$^{-10}$\\
    $\Sigma(res)^2$ & 9.97 $\times$ 10$^{-4}$\\
    \hline  
    $\dot{P}$ (day yr$^{-1}$) &  - (3.03 $\pm$ 0.26) $\times$ 10$^{-7}$\\
    $\dot{M_1}$ (M$_\odot$ yr$^{-1}$) & - (6.43 $\pm$ 1.35) $\times$ 10$^{-8}$\\
    $^\dagger\dot{M_1}$ (M$_\odot$ yr$^{-1}$) & - (2.97 $\pm$ 0.50) $\times$ 10$^{-8}$\\
    \hline
    \end{tabular}
    \begin{flushleft}
        {$^\dagger$: non-conservative transfer.}
    \end{flushleft}
    \label{tab:o-c}
\end{table}
\section{EVOLUTIONARY STATUS through numerical modeling}
\label{sec:evo}
To investigate the evolutionary nature of FASTT 1558, we compute various binary evolution models using Modules for Experiments in Stellar Astrophysics \citep[\texttt{MESA v25.12.1};][]{2011ApJS..192....3P, 2013ApJS..208....4P, 2015ApJS..220...15P, 2018ApJS..234...34P, 2019ApJS..243...10P, 2023ApJS..265...15J}. The initial conditions of a binary star at Zero Age Main Sequence (ZAMS) are crucial in determining its subsequent evolution and eventual fate. However, there is no unique, observationally established procedure for reconstructing the initial configuration of an observed contact binary. Semi-empirical approaches, such as those of \citet{2013MNRAS.430.2029Y, 2014MNRAS.437..185Y}, infer the progenitor masses and binary orbital period from present-day properties. But these prescriptions rely on simplified assumptions regarding mass transfer, mass and angular-momentum loss, and the onset of binary interaction, which may not adequately represent the evolutionary pathway of long-lived contact binaries and may overestimate the parameters at ZAMS. In this work, we therefore construct \texttt{MESA} models over a range of plausible initial binary configurations assuming conservative mass transfer. Given the observed total mass of FASTT 1558 (1.92 $\pm$ 0.31 $M_\odot$; see Table \ref{tab:params}), we consider total masses in the range \(1.6\leq M_{\rm tot}/M_\odot\leq2.2\). The initial primary mass \(M_{1,i}\) is varied from \(0.9\) to \(1.3\,M_\odot\) in steps of \(0.1\,M_\odot\), while the initial secondary mass \(M_{2,i}\) is chosen such that the total initial mass remains within the adopted range. We consider initial orbital periods in the range, \(P_i \in (0.5,0.9)\) days to obtain Case A interactions \citep{2025ApJ...995...19F}. We follow the evolutionary framework of \citet{2025ApJ...995...19F}, incorporating prescriptions for tidal deformation, energy transfer, and saturated magnetic braking, using their publicly available \texttt{MESA} implementation\footnote{DOI: \href{https://doi.org/10.5281/zenodo.17155744}{10.5281/zenodo.17155744}}. 

\begin{figure}[h]
    \centering
    \includegraphics[width=0.49\linewidth]{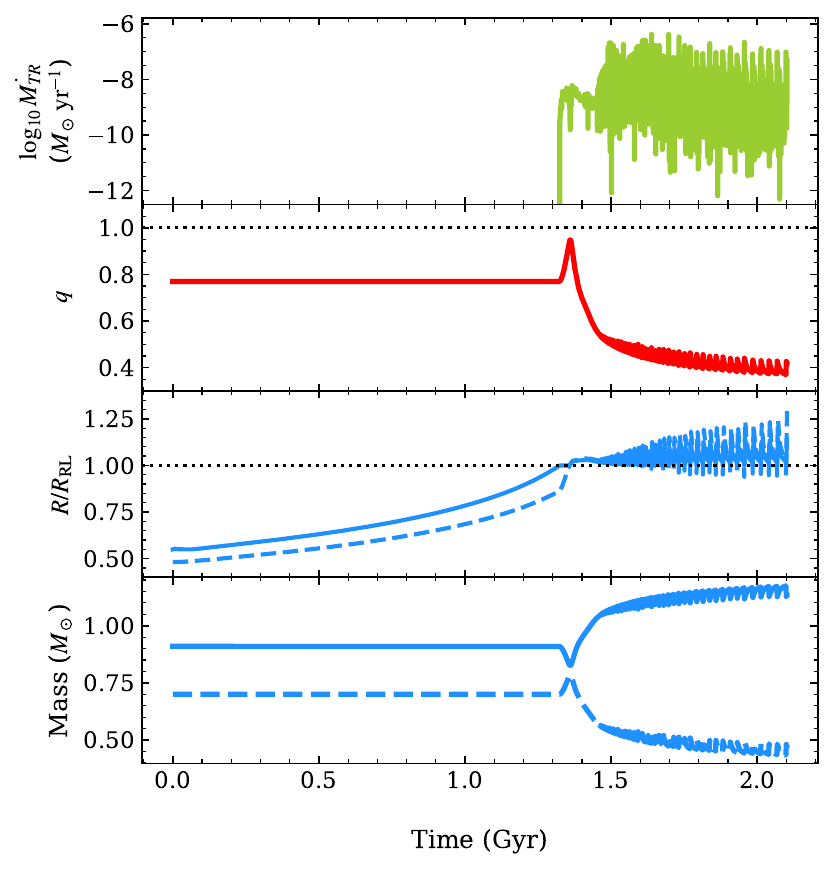}
    \hfill
    \includegraphics[width=0.49\linewidth]{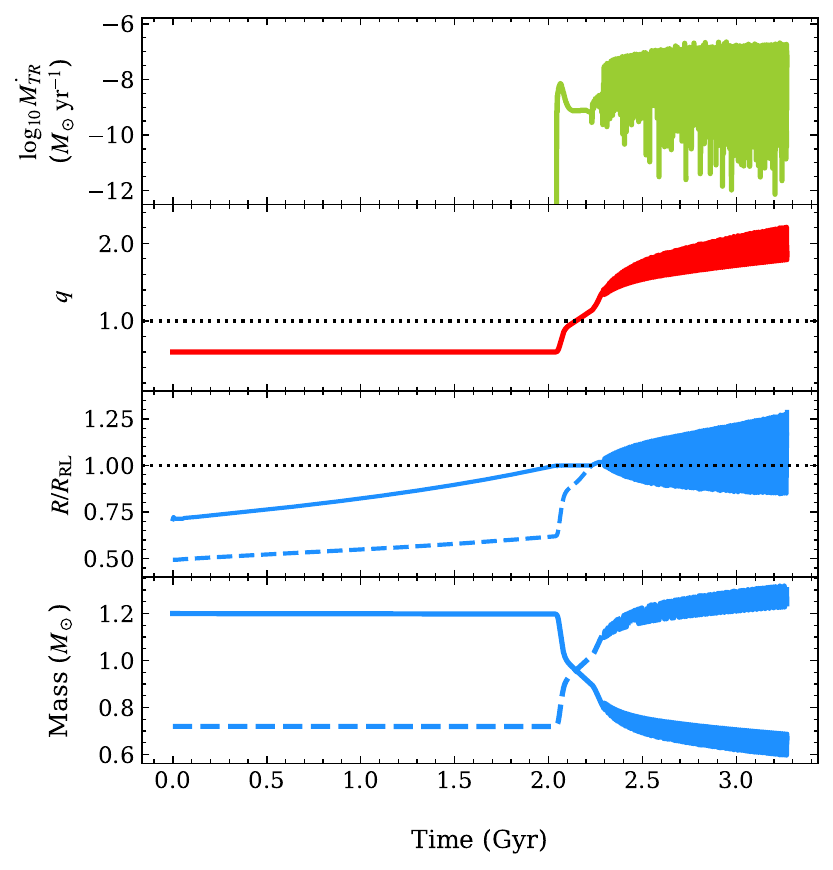}
    \caption{Mass-transfer rate, mass-ratio, radius and mass evolution models. In the third and fourth panels, the solid lines represent the primary components, while the dashed lines represent the secondary components. The black dotted horizontal lines in the second panels are references for $q = 1.0$ while in the third panels, the dotted lines indicate when both stars have $R/R_L \geq 1$, which is contact configuration. \textit{Left}: Model with $M_{1,i} = 0.91 \ M_{\odot}, \ M_{2,i} = 0.7 \ M_{\odot}, \ P_i = 0.67 \rm \ day$. \textit{Right}: Model with $M_{1,i} = 1.20 \ M_{\odot}, \ M_{2,i} = 0.72 \ M_{\odot}, \ P_i = 0.67 \rm \ day$. Both the models terminate due to \(L_2\) overflow.}
    \label{fig:tro_models}
\end{figure}
Our evolutionary grid suggests that the formation of low-mass-ratio contact binaries requires TROs. Furthermore, the amplitude of these oscillations must remain sufficiently small to prevent the system from reaching \(L_2\) overflow before attaining the observed low mass ratios. Although truncations are imposed after small number of TROs in previous studies, e.g., 20 TROs \citep{2025ApJ...995...19F}, we do not impose such a limit on the number of TROs. Instead, we allow the models to evolve through successive oscillations until they encounter \(L_2\) overflow. This allows us to investigate whether continued oscillations can drive the systems toward the mass ratios observed in systems such as FASTT 1558. Nevertheless, none of our models reaches \(q=0.15\), highlighting a limitation of the current evolutionary framework in reproducing the most extreme mass-ratio contact binaries. Figure \ref{fig:tro_models} illustrates two representative evolutionary sequences exhibiting different modes of TRO behaviour. In the first sequence (left panel), energy transfer from the donor causes the convective companion to expand substantially, leading to a reversal of the mass-transfer direction even before the mass ratio reaches \(q=1\). In the second sequence (right panel), mass ratio reversal occurs, where the initially more massive component loses sufficient mass through transfer to its companion, becoming the less massive star. Both sequences reach \(q\simeq0.4\) before terminating owing to \(L_2\) overflow.
\section{DISCUSSION AND CONCLUSION}
\label{sec:diss-con}
\begin{figure}[h]
    \centering
    \includegraphics[width=0.5\linewidth]{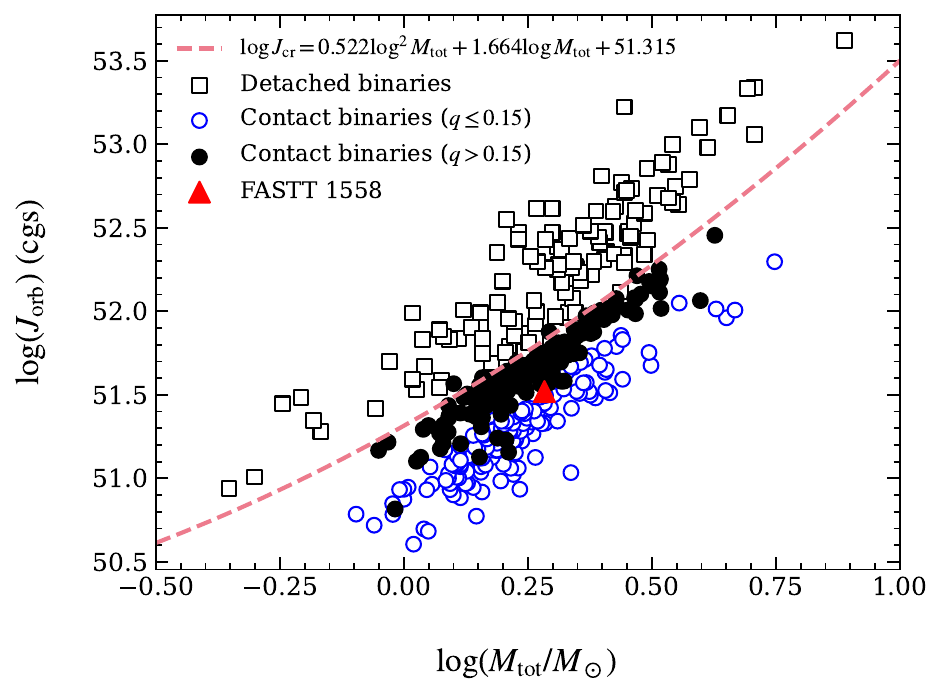}
    \caption{Location of FASTT 1558 in the diagram of \(\log J_{\rm orb}\) vs \(\log M_{\rm tot}\). Detached binaries are represented by open squares, while contact binaries with $q>0.15$ and $q\leq0.15$ are represented by filled circles and open circles, respectively. From our analysis, FASTT 1558 has $q=0.1544$.}    
    \label{fig:J_M}
\end{figure}

We present the first spectro-photometric and orbital period analysis of the totally eclipsing contact binary FASTT 1558. Photometric modeling with \texttt{PHOEBE} reveals that the system is a low mass-ratio contact binary with \(q = 0.1544 \pm 0.0013\) and a deep degree of contact, \(f=(61 \pm 1)\%\). We observe that the hotter component of binary is also the more massive one, classifying the system as an A-subtype contact binary. The location of FASTT 1558 in the \(J_{\rm orb}-M_{\rm tot}\) plane is shown in Figure \ref{fig:J_M}, together with known detached and contact binaries. The contact binary sample is taken from \citet{2022NewA...9101695Y} and \citet{2026AJ....171..254Y}, while the detached binary sample is from \citet{2006MNRAS.373.1483E}. The red dashed line represents the detached-contact binary boundary \citep{2006Ap&SS.304...75E}. FASTT 1558 (shown as a triangle) lies close to the boundary separating contact binaries with \(q\leq0.15\) from those with \(q>0.15\), placing it among the systems with the lowest observed mass ratios. We note that this boundary is an empirical threshold in the mass-ratio, used to distinguish two observational populations of contact binaries, and not a theoretical critical limit associated with Darwin instability.

We investigate the dynamical stability of the system based on the ratio of spin to orbital angular momentum,
\begin{equation}
    \frac{J_{spin}}{J_{orbit}} = \frac{1+q}{q}[(k_1r_1)^2+ (k_2r_2)^2q].
\end{equation}
The $J_{spin}/J_{orbit}$ value of 0.15 $\pm$ 0.02 \citep[$< \frac{1}{3}$, Darwin stable;][]{darwin1879viii, 1980A&A....92..167H} suggests that the system is dynamically stable. Here, the values of $k_1^2$ and $k_2^2$ are taken as 0.06 and 0.10, respectively, as suggested by \citet{2026A&A...710A..49P} for extremely low mass ratio contact binaries to account for strong Roche distortion of the secondaries.

Our orbital period analysis indicates a decreasing period of (-3.03 $\pm$ 0.26) $\times$ 10$^{-7}$ day yr$^{-1}$. The period decrease may be due to magnetic braking (MB) and associated angular momentum loss, evaluated using the following formulation \citep{1994ASPC...56..228B, 2026AJ....171..347P},
\begin{equation}
    \frac{dP}{dt} \approx - 1.1 \times 10^{-8}\frac{(1+q)^2}{q}(M_1+M_2)^{-\frac{5}{3}} k^2 (M_1R_1^4 - M_2R_2^4)P^{-\frac{7}{3}}.
\end{equation}
Using gyration $k^2$ as 0.06 \citep{2006MNRAS.369.2001L} and other parameter values from our analysis, we get $\dot{P} = (-1.92 \pm 0.63) \times 10^{-7}$ day yr$^{-1}$. As discussed in \citet{2025ApJ...995...19F}, for low-mass contact binaries magnetic braking is perhaps saturated and therefore the derived value of $\dot{P}$ would be an upper limit for the period decay rate, as the method of \citet{1994ASPC...56..228B} does not consider the effects of saturated magnetic braking. Thus, MB decreases the orbital period at a rate smaller than the observed period-decrease rate. An alternative mechanism is angular momentum loss driven by gravitational wave radiation (GWR). As suggested by \citet{2012A&A...542A.124L}, combining expressions from \citet{2010eea..book.....K} and \citet{2001icbs.book.....H},
\begin{equation}
\frac{\dot{P}}{P} = 3 \frac{\dot{J}}{J} = -1.27 \times 10^{-8} \ \frac{M_1 M_2}{(M_1 + M_2)^{1/3} \, M_\odot^{5/3}} 
\left( \frac{P_{\mathrm{orb}}}{\mathrm{h}} \right)^{-8/3} \ \mathrm{yr}^{-1}
\end{equation}
we get $\dot{P} = (-3.13 \pm 0.59) \times 10^{-12}$ day yr$^{-1}$, which is nearly 5 orders of magnitude lower than what we observe. Hence we expect no significant contribution from MB and GWR.

To explain the observed period decrease rate, we examine the mechanisms of mass transfer from the primary to the secondary star and/or mass (and angular momentum) loss from the system through outer Lagrange point, L$_2$. Using the formulations given by \citet{2001icbs.book.....H} and \citet{1958BAN....14..131K} for conservative and non-conservative mass loss respectively,
\begin{equation}
    \frac{\dot{P}}{P} = \frac{3 \dot{M}_1 (M_1 - M_2)}{M_1 M_2}
\end{equation}

\begin{equation}
    \frac{\dot{P}}{P} = 3 \dot{M}_1 \left[\frac{(M_1 + M_2)}{M_1 M_2} \frac{d^2}{a^2} - \frac{M_2}{M_1 (M_1 + M_2)}\right]
\end{equation}
where $d$ is the distance from the binary centre of mass to $L_2$, we estimate the mass transfer rate $\dot{M}_1$ required to reproduce the observed $\dot{P}$ values, given in Table \ref{tab:o-c}. The thermal timescale of the primary star can be measured as, $\tau_{th} = \frac{GM_1^2}{R_1L_1}$ = (7.83 $\pm$ 2.67) $\times$ 10$^6$ years. Thus, the thermal mass transfer rate, $M_1/\tau_{th}$ = (2.12 $\pm$ 0.43) $\times$ 10$^{-7}$ M$_\odot$ yr$^{-1}$, is $\sim$ 3.33 times the conservative mass transfer rate, suggesting that the mass transfer is likely occurring on the thermal (Kelvin–Helmholtz) timescale.

From ADFOSC spectra of FASTT 1558, we determine the atmospheric parameters of the binary. These are fairly consistent with the previously reported LAMOST and Gaia DR3 values. For the first time, we investigate the chromospheric activity of the system through the analysis of the well known chromospheric diagnostic lines H$\alpha$ and Ca\textsc{ii}. The temporal behaviour of the EW of H$\alpha$ and Ca\textsc{ii} lines presented in Figure \ref{fig:figure-ew}, reveals a dynamically evolving line-forming region. The observed variations may be caused by temperature variations, pressure or density effects and surface inhomogeneities (spots or plages) of the binary star \citep{2006A&A...446.1129B,2020A&A...635A..89M}. The \texttt{PHOEBE} light-curve models do not require spot modelling, suggesting that the observed spectroscopic variability may not be attributed solely to large-scale photospheric spots. The presence of an inverse P Cygni profile in the chromospheric H$\alpha$ lines in all observations, may suggest the transfer of mass along the line of sight \citep{2008A&A...487..709Z}. However, the observed feature alone does not establish infall as its definitive physical origin. Given the limited spectral resolution and the absence of higher-resolution spectroscopic observations, this interpretation remains tentative and requires further investigation. We note that, the systematic uncertainties associated with the spectral subtraction procedure, including template mismatch, continuum normalisation, instrumental-resolution matching and residual telluric contamination, may also contribute to the observed profile. We do not expect to reliably detect wavelength shifts in such low resolution spectra (measured displacements are generally below one instrumental resolution element). Therefore, the variations in the modelled shifts of the redshifted component across the 36 ADFOSC spectra (See Table \ref{tab:p-ew}) may also indicate uncertainties in the spectral-subtraction procedure. Moreover, both the emission and absorption components of the profile exhibit time variability in their equivalent widths, indicating a dynamical chromosphere.
\begin{figure}[h]
    \centering
    \includegraphics[width=0.5\linewidth]{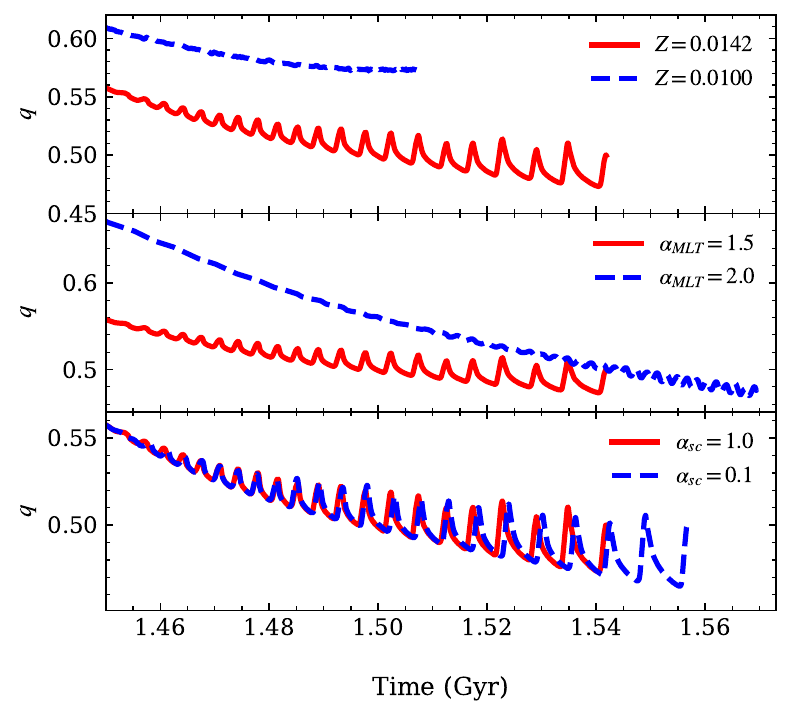}
    \caption{Mass-ratio evolution under different prescriptions for metallicity and microphysics with $M_{1,i} = 0.91 \ M_{\odot},$ $\ M_{2,i} = 0.7 \ M_{\odot}, \ P_i = 0.67 \rm \ day$. The model with lower metallicity, Z (top) terminates due to \(L_2\) overflow. The model with higher mixing length parameter, \(\alpha_{MLT}\) (middle) starts TROs slightly after the default model in solid red, while the model with lower semiconvection efficiency parameter, \(\alpha_{sc}\) (bottom) does not show such delay.}
    \label{fig:q-comp}
\end{figure}

We computed a grid of binary models in \texttt{MESA} spanning a range of initial masses and orbital periods using the prescriptions for tidal deformation, mass and energy transfer, and saturated magnetic braking. Our evolutionary calculations are consistent with the proposed role of TROs in driving the secular evolution of contact binaries toward progressively smaller mass ratios. However, none of the models evolve to the low mass ratio of \(q\sim0.15\) inferred for the observed system; instead, the models terminate when the donor reaches \(L_2\) overflow. To investigate the sensitivity of the evolutionary pathway to the adopted physical assumptions, we additionally explored the effects of metallicity and the treatment of convection and semi-convection as illustrated in Figure \ref{fig:q-comp}. The solid red curves correspond to the default prescriptions of \citet{2025ApJ...995...19F}, which are adopted in our calculations, while the dashed blue curves show representative models obtained by modifying individual parameters. Although these changes somewhat alter the evolutionary behaviour, the mass ratio does not decrease rapidly over the timescales covered by our calculations. Owing to the high computational cost of the TRO phase, the termination criterion for these additional models presented in Figure \ref{fig:q-comp}, is adopted to be 20 TROs, while the main evolutionary grid of Section \ref{sec:evo} is followed until \(L_2\) overflow. Therefore, we cannot exclude the possibility that such combinations of initial binary parameters or different physical prescriptions may produce evolutionary pathways that reach the observed low mass ratio. A more extensive exploration of the relevant parameter space, including the prescriptions governing rotation and stellar mixing, is beyond the scope of the present study and will be investigated separately.

The evolution of extremely low mass-ratio contact binaries provides a stringent test of binary-evolution models. Systems such as FASTT 1558, serve as an important observational benchmark for testing the physical prescriptions governing contact-binary evolution. Continued efforts to reconcile evolutionary models with such extreme systems will be essential for establishing how contact binaries attain such low mass ratios and ultimately approach merger.
\begin{acknowledgments}
We sincerely thank the anonymous referee for the constructive comments and suggestions, which have helped us improve the clarity and quality of the manuscript. A.B. would like to thank Dr. Matthias Fabry for his insightful comments and valuable support throughout this study on FASTT 1558. A.B. acknowledges the Department of Science and Technology (DST), Government of India, for the award of the INSPIRE fellowship (IF230384), and the research facilities provided by the Department of Physics, IIT Hyderabad. 
\end{acknowledgments}

\facilities{TESS, ASAS-SN, DOT:3.6m(ADFOSC)}
\software{IRAF \citep{1986SPIE..627..733T}, astropy \citep{2013A&A...558A..33A}, scipy \citep{2020NaMet..17..261V}, lightkurve \citep{2018ascl.soft12013L}, ULySS \citep{2009A&A...501.1269K}, iSTARMOD \citep{2026AJ....171...15L}, PHOEBE \citep{2005ApJ...628..426P}, MESA \citep{2011ApJS..192....3P}. We utilized the language model ChatGPT-5 to assist in restructuring and refining a few sentences in the manuscript.}

\begin{contribution}
\textbf{A.B.}: Writing – original draft, Visualization, Methodology, Investigation, Formal analysis, Conceptualization. \textbf{M. P.}: Writing – review \& editing, Supervision, Resources, Project administration. \textbf{K. P.}: Investigation - observation, Revision of Manuscript. \textbf{N. M. T.}: Visualization, Investigation. \textbf{K. M.}: Supervision, Revision of Manuscript. \textbf{I.T.}: Investigation.
\end{contribution}

\bibliographystyle{aasjournalv7}

\appendix

\begin{table}[h]
\centering
\caption{Measured equivalent widths of H$\alpha$ and Ca\textsc{ii} lines for FASTT 1558. Observations were taken on November 16, 2025. $\Delta\lambda_{\rm H\alpha,\rm{chrom}}$ is the displacement of the redshifted component from the H$\alpha$ rest-frame line centre at 6562.8 \AA, in chromospheric spectra. The errors are quoted with 1$\sigma$ uncertainty.}
\begin{tabular}{cccccc}
\hline
Time & BJD & Phase & H$\alpha$ & Ca \textsc{ii} & $\Delta\lambda_{\rm H\alpha,\rm{chrom}}$\\
(IST)&     &       & EW (\AA)  & EW (\AA) & (\AA)\\
\hline
\hline
19:17:48.850 & 2460996.077236 & 0.511038 & 3.77 $\pm$ 0.04 & 1.92 $\pm$ 0.06 & 7.73 $\pm$ 0.12\\
19:28:25.196 & 2460996.084601 & 0.526467 & 3.64 $\pm$ 0.04 & 1.72 $\pm$ 0.05 & 7.65 $\pm$ 0.11\\
19:39:01.719 & 2460996.091967 & 0.541900 & 3.60 $\pm$ 0.04 & 1.83 $\pm$ 0.05 & 4.34 $\pm$ 0.07\\
19:49:38.064 & 2460996.099332 & 0.557330 & 3.55 $\pm$ 0.04 & 1.90 $\pm$ 0.07 & 7.88 $\pm$ 0.10\\
20:00:48.229 & 2460996.107087 & 0.573579 & 3.59 $\pm$ 0.05 & 1.68 $\pm$ 0.06 & 8.09 $\pm$ 0.11\\
20:09:24.565 & 2460996.113063 & 0.586098 & 3.68 $\pm$ 0.05 & 1.78 $\pm$ 0.07 & 8.16 $\pm$ 0.11\\
20:18:00.914 & 2460996.119039 & 0.598618 & 3.69 $\pm$ 0.05 & 1.61 $\pm$ 0.06 & 8.00 $\pm$ 0.11\\
20:26:37.241 & 2460996.125014 & 0.611137 & 3.58 $\pm$ 0.05 & 1.96 $\pm$ 0.08 & 7.94 $\pm$ 0.12\\
20:35:13.590 & 2460996.130990 & 0.623656 & 3.55 $\pm$ 0.05 & 2.04 $\pm$ 0.09 & 7.93 $\pm$ 0.12\\
20:43:49.938 & 2460996.136965 & 0.636176 & 3.47 $\pm$ 0.05 & 2.09 $\pm$ 0.07 & 8.19 $\pm$ 0.12\\
20:52:26.455 & 2460996.142943 & 0.648700 & 3.88 $\pm$ 0.04 & 1.90 $\pm$ 0.08 & 8.25 $\pm$ 0.10\\
21:01:02.806 & 2460996.148919 & 0.661219 & 3.66 $\pm$ 0.04 & 2.30 $\pm$ 0.09 & 8.19 $\pm$ 0.11\\
21:10:15.396 & 2460996.155314 & 0.674618 & 3.64 $\pm$ 0.05 & 2.42 $\pm$ 0.12 & 8.27 $\pm$ 0.11\\
21:15:51.737 & 2460996.159206 & 0.682773 & 3.65 $\pm$ 0.06 & 2.70 $\pm$ 0.13 & 8.47 $\pm$ 0.10\\
21:21:28.087 & 2460996.163099 & 0.690928 & 3.69 $\pm$ 0.05 & 2.94 $\pm$ 0.22 & 5.37 $\pm$ 0.06\\
21:27:04.432 & 2460996.166991 & 0.699083 & 3.55 $\pm$ 0.06 & 2.49 $\pm$ 0.13 & 5.46 $\pm$ 0.06\\
21:31:40.803 & 2460996.170190 & 0.705784 & 3.70 $\pm$ 0.05 & 3.63 $\pm$ 0.28 & 5.37 $\pm$ 0.06\\
21:38:17.137 & 2460996.174777 & 0.715394 & 3.63 $\pm$ 0.05 & 3.93 $\pm$ 0.28 & 8.42 $\pm$ 0.08\\
21:43:47.953 & 2460996.178605 & 0.723415 & 3.59 $\pm$ 0.05 & 2.92 $\pm$ 0.15 & 8.15 $\pm$ 0.08\\
21:49:29.812 & 2460996.182561 & 0.731704 & 3.68 $\pm$ 0.06 & 2.85 $\pm$ 0.16 & 8.36 $\pm$ 0.09\\
21:55:06.138 & 2460996.186454 & 0.739859 & 3.57 $\pm$ 0.05 & 2.20 $\pm$ 0.10 & 5.37 $\pm$ 0.06\\
22:00:42.471 & 2460996.190346 & 0.748014 & 3.64 $\pm$ 0.06 & 2.99 $\pm$ 0.19 & 8.28 $\pm$ 0.06\\
22:06:18.816 & 2460996.194239 & 0.756169 & 3.65 $\pm$ 0.06 & 2.67 $\pm$ 0.16 & 8.36 $\pm$ 0.06\\
22:15:16.671 & 2460996.200463 & 0.769210 & 3.61 $\pm$ 0.06 & 2.16 $\pm$ 0.12 & 9.01 $\pm$ 0.08\\
22:20:53.022 & 2460996.204356 & 0.777365 & 3.61 $\pm$ 0.06 & 2.13 $\pm$ 0.12 & 7.55 $\pm$ 0.08\\
22:26:29.337 & 2460996.208248 & 0.785520 & 3.77 $\pm$ 0.06 & 2.34 $\pm$ 0.12 & 6.71 $\pm$ 0.06\\
22:32:05.696 & 2460996.212141 & 0.793675 & 3.80 $\pm$ 0.06 & 3.23 $\pm$ 0.25 & 7.65 $\pm$ 0.06\\
22:37:42.026 & 2460996.216033 & 0.801830 & 3.70 $\pm$ 0.06 & 2.88 $\pm$ 0.16 & 6.95 $\pm$ 0.08\\
22:43:18.352 & 2460996.219925 & 0.809985 & 3.56 $\pm$ 0.06 & 3.08 $\pm$ 0.20 & 6.92 $\pm$ 0.09\\
22:48:54.687 & 2460996.223818 & 0.818140 & 3.82 $\pm$ 0.06 & 3.53 $\pm$ 0.25 & 6.80 $\pm$ 0.07\\
22:55:36.753 & 2460996.228471 & 0.827888 & 3.68 $\pm$ 0.06 & 2.60 $\pm$ 0.12 & 7.44 $\pm$ 0.07\\
23:01:13.094 & 2460996.232363 & 0.836043 & 3.62 $\pm$ 0.06 & 2.47 $\pm$ 0.13 & 6.70 $\pm$ 0.07\\
23:06:49.433 & 2460996.236256 & 0.844198 & 3.71 $\pm$ 0.06 & 2.31 $\pm$ 0.13 & 6.75 $\pm$ 0.06\\
23:12:25.753 & 2460996.240148 & 0.852353 & 3.64 $\pm$ 0.06 & 2.69 $\pm$ 0.14 & 7.17 $\pm$ 0.09\\
23:23:42.025 & 2460996.247974 & 0.868750 & 3.74 $\pm$ 0.06 & 2.26 $\pm$ 0.11 & 6.81 $\pm$ 0.07\\
23:29:58.039 & 2460996.252326 & 0.877867 & 3.61 $\pm$ 0.06 & 2.39 $\pm$ 0.11 & 6.80 $\pm$ 0.10\\
\hline
\end{tabular}
\label{tab:p-ew}
\end{table}

\begin{figure}[h]
    \centering
    \includegraphics[width=\linewidth]{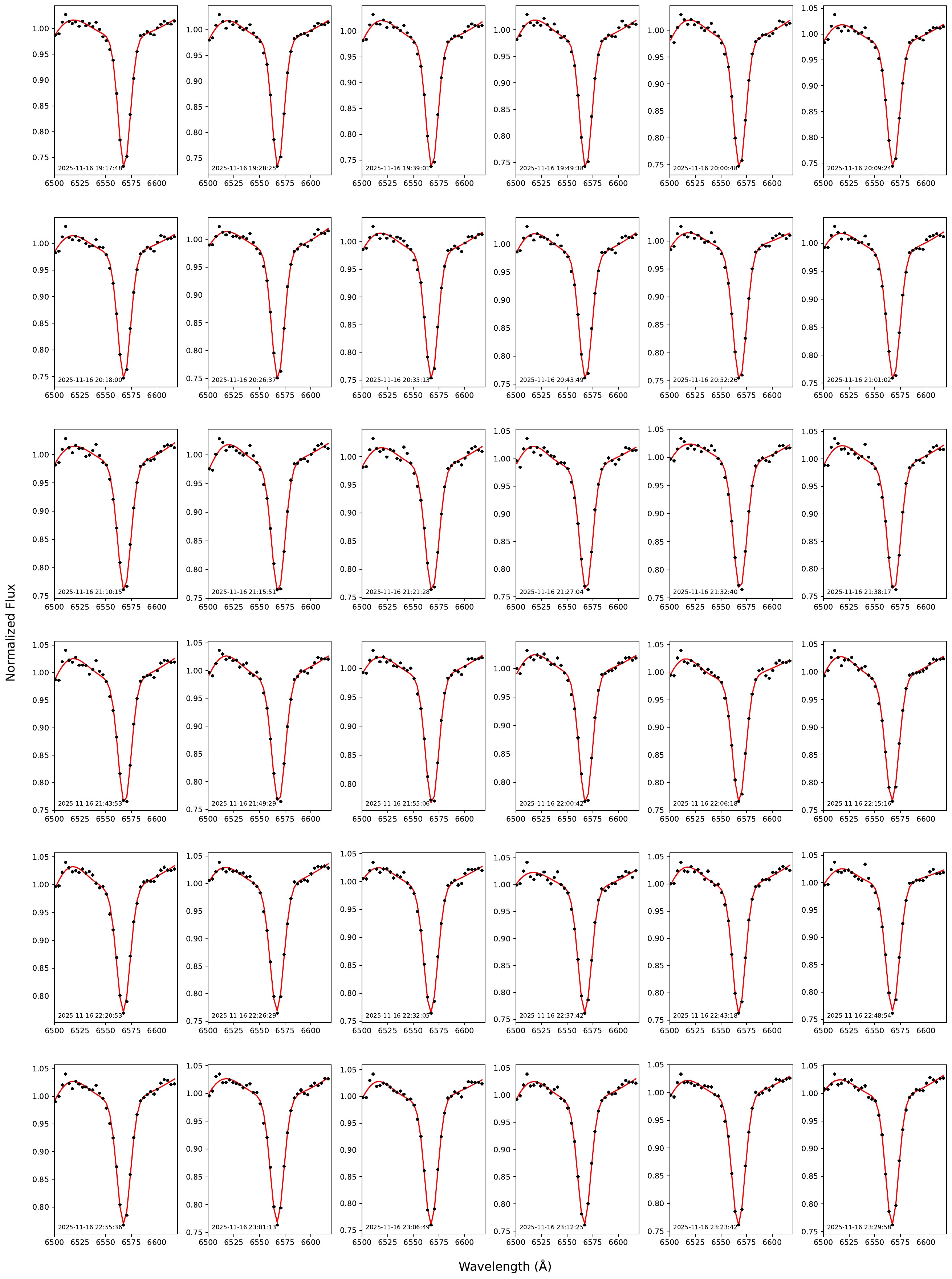}
    \caption{H$\alpha$ absorption lines with fitted model.}
    \label{fig:h-a}
\end{figure}

\begin{figure}[h]
    \centering
    \includegraphics[width=\linewidth]{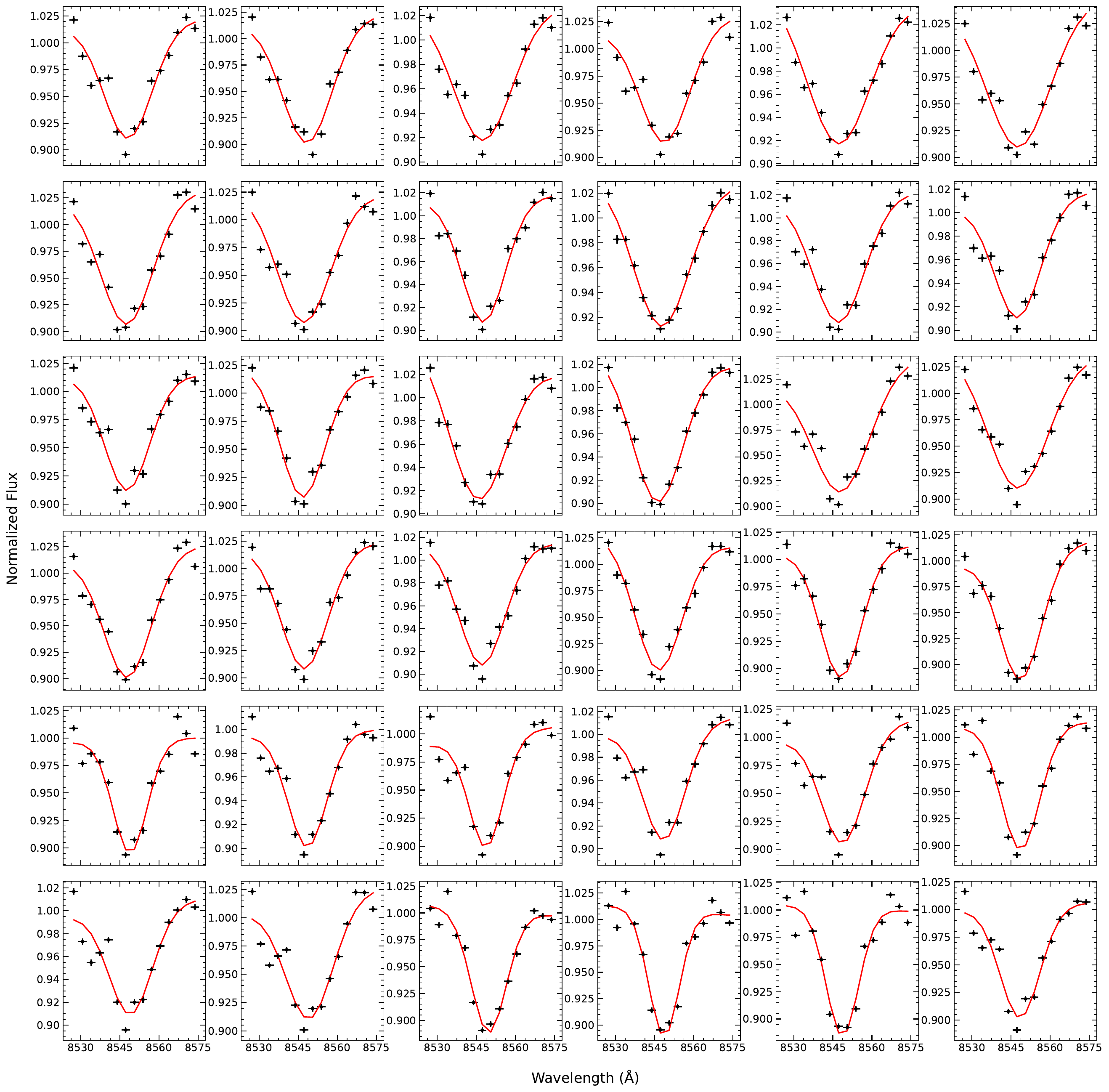}
    \caption{Ca\textsc{ii} absorption lines with fitted model.}
    \label{fig:ca-ii}
\end{figure}
\begin{table}[h]
    \centering
    \caption{Times of primary minima and O-C values for FASTT 1558. The minima are reported after subtracting 2,400,000.}
    \begin{tabular}{lccc||cccc}
    \hline
Eclipse Timing & Error & O-C & Epoch No. & Eclipse Timing & Error & O-C & Epoch No.\\
    (BJD) & (BJD) &  (days)  & (days) & (BJD) & (BJD) & (days)   & (days)\\
    \hline
    \hline
$^*$56534.5058 & 0.0017 & -0.0004 & 3637 & 60224.5156 & 0.0006 & -0.0356 & 11368 \\
$^*$56785.0843 & 0.0027 & -0.0071 & 4162 & 60224.9907 & 0.0009 & -0.0378 & 11369 \\
$^*$56883.8860 & 0.0011 & -0.0083 & 4369 & 60225.4682 & 0.0010 & -0.0377 & 11370 \\
$^*$57185.0657 & 0.0012 & -0.0080 & 5000 & 60225.9438 & 0.0008 & -0.0393 & 11371 \\
$^*$57359.7616 & 0.0011 & -0.0057 & 5366 & 60226.4214 & 0.0006 & -0.0390 & 11372 \\
$^*$57499.1292 & 0.0018 & -0.0110 & 5658 & 60226.9011 & 0.0009 & -0.0367 & 11373 \\
$^*$57639.9284 & 0.0011 & -0.0169 & 5953 & 60227.3753 & 0.0009 & -0.0397 & 11374 \\
$^*$57941.1041 & 0.0012 & -0.0208 & 6584 & 60227.8566 & 0.0008 & -0.0358 & 11375 \\
$^*$58310.0558 & 0.0025 & -0.0259 & 7357 & 60228.3326 & 0.0010 & -0.0371 & 11376 \\
$^*$58430.8128 & 0.0016 & -0.0270 & 7610 & 60228.8131 & 0.0005 & -0.0339 & 11377 \\
$^\dagger$53637.7419 & - & -0.0002   & -2432 & 60229.2862 & 0.0008 & -0.0381 & 11378 \\
$^\dagger$54798.5479 & - & 0          & 0    & 60229.7627 & 0.0012 & -0.0389 & 11379 \\
59447.9472 & 0.0001 & -0.0287 & 9741 & 60230.2401 & 0.0010 & -0.0388 & 11380 \\
59448.4247 & 0.0001 & -0.0286 & 9742 & 60230.7174 & 0.0007 & -0.0388 & 11381 \\
59448.9018 & 0.0001 & -0.0287 & 9743 & 60231.1955 & 0.0007 & -0.0379 & 11382 \\
59449.3792 & 0.0001 & -0.0286 & 9744 & 60231.6725 & 0.0007 & -0.0383 & 11383 \\
59449.8565 & 0.0001 & -0.0287 & 9745 & 60232.1495 & 0.0008 & -0.0386 & 11384 \\
59450.3334 & 0.0001 & -0.0291 & 9746 & 60232.6271 & 0.0010 & -0.0383 & 11385 \\
59450.8108 & 0.0001 & -0.0289 & 9747 & 60533.8028 & 0.0008 & -0.0421 & 12016 \\
59451.2881 & 0.0001 & -0.0290 & 9748 & 60534.2798 & 0.0007 & -0.0424 & 12017 \\
59461.7884 & 0.0001 & -0.0294 & 9770 & 60534.7573 & 0.0009 & -0.0422 & 12018 \\
59462.2660 & 0.0001 & -0.0291 & 9771 & 60535.2335 & 0.0009 & -0.0433 & 12019 \\
59462.7433 & 0.0001 & -0.0291 & 9772 & 60535.7111 & 0.0007 & -0.0430 & 12020 \\
59463.2207 & 0.0001 & -0.0289 & 9773 & 60536.1892 & 0.0008 & -0.0422 & 12021 \\
59463.6978 & 0.0001 & -0.0292 & 9774 & 60536.6671 & 0.0009 & -0.0416 & 12022 \\
59464.1753 & 0.0001 & -0.0290 & 9775 & 60538.5740 & 0.0005 & -0.0439 & 12026 \\
59464.6526 & 0.0001 & -0.0290 & 9776 & 60539.0546 & 0.0008 & -0.0406 & 12027 \\
59465.1299 & 0.0001 & -0.0290 & 9777 & 60539.5297 & 0.0008 & -0.0428 & 12028 \\
59465.6071 & 0.0001 & -0.0292 & 9778 & 60540.0080 & 0.0009 & -0.0418 & 12029 \\
59466.0847 & 0.0001 & -0.0289 & 9779 & 60540.4860 & 0.0007 & -0.0411 & 12030 \\
59466.5618 & 0.0001 & -0.0291 & 9780 & 60540.9617 & 0.0007 & -0.0428 & 12031 \\
59467.0393 & 0.0001 & -0.0289 & 9781 & 60541.4395 & 0.0011 & -0.0423 & 12032 \\
59467.5164 & 0.0001 & -0.0290 & 9782 & 60542.3929 & 0.0008 & -0.0435 & 12034 \\
59467.9932 & 0.0001 & -0.0296 & 9783 & 60542.8717 & 0.0013 & -0.0420 & 12035 \\
59468.4709 & 0.0001 & -0.0291 & 9784 & 60544.7817 & 0.0012 & -0.0412 & 12039 \\
59468.9483 & 0.0001 & -0.0291 & 9785 & 60545.7363 & 0.0007 & -0.0412 & 12041 \\
59469.4256 & 0.0001 & -0.0291 & 9786 & 60546.2121 & 0.0009 & -0.0427 & 12042 \\
59469.9027 & 0.0001 & -0.0293 & 9787 & 60546.6932 & 0.0005 & -0.0389 & 12043 \\
59470.3800 & 0.0001 & -0.0293 & 9788 & 60547.1657 & 0.0009 & -0.0436 & 12044 \\
60209.2380 & 0.0010 & -0.0394 & 11336 & 60547.6447 & 0.0009 & -0.0420 & 12045 \\
60210.1928 & 0.0006 & -0.0392 & 11338 & 60548.1221 & 0.0009 & -0.0419 & 12046 \\
60210.6684 & 0.0008 & -0.0409 & 11339 & 60549.0771 & 0.0011 & -0.0416 & 12048 \\
60211.1476 & 0.0008 & -0.0391 & 11340 & 60549.5533 & 0.0009 & -0.0427 & 12049 \\
60211.6244 & 0.0013 & -0.0395 & 11341 & 60550.0317 & 0.0008 & -0.0415 & 12050 \\
60212.1024 & 0.0008 & -0.0388 & 11342 & 60550.5082 & 0.0008 & -0.0423 & 12051 \\
60212.5805 & 0.0009 & -0.0380 & 11343 & 60550.9828 & 0.0005 & -0.0451 & 12052 \\
60213.0576 & 0.0008 & -0.0382 & 11344 & 60551.4615 & 0.0005 & -0.0437 & 12053 \\
60213.5349 & 0.0008 & -0.0383 & 11345 & 60551.9402 & 0.0009 & -0.0422 & 12054 \\
    \hline
    \end{tabular}
    \label{tab:o-c1}
\end{table}
\begin{table}[]
    \centering
    \begin{tabular}{cccc||cccc}
    \hline
Eclipse Timing & Error & O-C & Epoch No. & Eclipse Timing & Error & O-C & Epoch No.\\
    (BJD) & (BJD) &  (days)  & (days) & (BJD) & (BJD) & (days)   & (days)\\
    \hline
    \hline
60214.0118 & 0.0008 & -0.0387 & 11346 & 60552.4189 & 0.0013 & -0.0409 & 12055 \\
60214.4905 & 0.0011 & -0.0373 & 11347 & 60553.3717 & 0.0009 & -0.0427 & 12057 \\
60214.9662 & 0.0014 & -0.0389 & 11348 & 60553.8491 & 0.0009 & -0.0426 & 12058 \\
60215.4437 & 0.0008 & -0.0387 & 11349 & 60554.3262 & 0.0008 & -0.0428 & 12059 \\
60215.9205 & 0.0006 & -0.0392 & 11350 & 60555.2817 & 0.0009 & -0.0419 & 12061 \\
60216.3981 & 0.0009 & -0.0389 & 11351 & 60555.7584 & 0.0007 & -0.0426 & 12062 \\
60216.8761 & 0.0008 & -0.0382 & 11352 & 60556.2381 & 0.0013 & -0.0401 & 12063 \\
60217.3527 & 0.0010 & -0.0389 & 11353 & 60556.7141 & 0.0011 & -0.0414 & 12064 \\
60222.1255 & 0.0007 & -0.0391 & 11363 & 60557.1917 & 0.0008 & -0.0411 & 12065 \\
60222.6034 & 0.0009 & -0.0386 & 11364 & 60557.6647 & 0.0024 & -0.0455 & 12066 \\
60223.5591 & 0.0010 & -0.0375 & 11366 & 60558.1466 & 0.0009 & -0.0408 & 12067 \\
60224.0363 & 0.0009 & -0.0376 & 11367 & 60559.0992 & 0.0008 & -0.0428 & 12069 \\
    \hline
    \end{tabular}
    \begin{flushleft}
        $^*$: ASAS-SN data. $^\dagger$: VarAstro data. All the other data are from TESS.
    \end{flushleft}    
\end{table}
\begin{table}[]
    \centering
    \caption{Times of secondary minima and O-C values for FASTT 1558. The minima are reported after subtracting 2,400,000.}
    \begin{tabular}{cccc||cccc}
    \hline
Eclipse Timing & Error & O-C & Epoch No. & Eclipse Timing & Error & O-C & Epoch No.\\
    (BJD) & (BJD) &  (days)  & (days) & (BJD) & (BJD) & (days)   & (days)\\
    \hline
    \hline
$^\dagger$54798.7869 & - & 0 & 0 & 60228.0919 & 0.0008 & -0.0395 & 11375 \\
$^\dagger$55864.6069 & - & 0.0021 & 2233 & 60228.5744 & 0.0022 & -0.0343 & 11376 \\
59448.1858 & 0.0001 & -0.0292 & 9741 & 60229.0486 & 0.0008 & -0.0374 & 11377 \\
59448.6631 & 0.0001 & -0.0292 & 9742 & 60229.5311 & 0.0003 & -0.0322 & 11378 \\
59449.1406 & 0.0001 & -0.0290 & 9743 & 60230.0039 & 0.0013 & -0.0366 & 11379 \\
59449.6178 & 0.0001 & -0.0291 & 9744 & 60230.4803 & 0.0011 & -0.0376 & 11380 \\
59450.5724 & 0.0001 & -0.0291 & 9746 & 60230.9515 & 0.0004 & -0.0436 & 11381 \\
59451.0499 & 0.0001 & -0.0289 & 9747 & 60231.4403 & 0.0004 & -0.0323 & 11382 \\
59451.5269 & 0.0001 & -0.0292 & 9748 & 60231.9064 & 0.0004 & -0.0434 & 11383 \\
59461.5504 & 0.0001 & -0.0291 & 9769 & 60232.3848 & 0.0009 & -0.0422 & 11384 \\
59462.0276 & 0.0001 & -0.0292 & 9770 & 60534.0432 & 0.0014 & -0.0406 & 12016 \\
59462.5050 & 0.0001 & -0.0292 & 9771 & 60534.5196 & 0.0051 & -0.0416 & 12017 \\
59462.9821 & 0.0001 & -0.0293 & 9772 & 60535.0001 & 0.0069 & -0.0383 & 12018 \\
59463.4596 & 0.0001 & -0.0291 & 9773 & 60535.4798 & 0.0005 & -0.0359 & 12019 \\
59463.9368 & 0.0001 & -0.0292 & 9774 & 60535.9552 & 0.0008 & -0.0379 & 12020 \\
59464.4142 & 0.0001 & -0.0291 & 9775 & 60536.9020 & 0.0006 & -0.0457 & 12022 \\
59464.8914 & 0.0001 & -0.0293 & 9776 & 60537.8593 & 0.0008 & -0.0430 & 12024 \\
59465.3689 & 0.0001 & -0.0291 & 9777 & 60538.3433 & 0.0005 & -0.0363 & 12025 \\
59465.8461 & 0.0001 & -0.0292 & 9778 & 60538.8191 & 0.0005 & -0.0378 & 12026 \\
59466.3236 & 0.0001 & -0.0289 & 9779 & 60539.7681 & 0.0014 & -0.0433 & 12028 \\
59466.8007 & 0.0001 & -0.0291 & 9780 & 60540.7301 & 0.0005 & -0.0360 & 12030 \\
59467.2778 & 0.0001 & -0.0294 & 9781 & 60541.2041 & 0.0007 & -0.0393 & 12031 \\
59467.7553 & 0.0001 & -0.0292 & 9782 & 60541.6793 & 0.0009 & -0.0414 & 12032 \\
59468.2324 & 0.0001 & -0.0294 & 9783 & 60542.6320 & 0.0022 & -0.0433 & 12034 \\
59468.7099 & 0.0001 & -0.0292 & 9784 & 60543.1079 & 0.0006 & -0.0447 & 12035 \\
59469.1871 & 0.0001 & -0.0293 & 9785 & 60545.0202 & 0.0010 & -0.0417 & 12039 \\
59469.6646 & 0.0001 & -0.0291 & 9786 & 60545.9732 & 0.0019 & -0.0433 & 12041 \\
59470.1418 & 0.0001 & -0.0292 & 9787 & 60546.4523 & 0.0008 & -0.0415 & 12042 \\
60209.4779 & 0.0011 & -0.0385 & 11336 & 60546.9268 & 0.0034 & -0.0443 & 12043 \\
60210.9089 & 0.0012 & -0.0395 & 11339 & 60547.4078 & 0.0009 & -0.0406 & 12044 \\
60211.8664 & 0.0008 & -0.0366 & 11341 & 60547.8866 & 0.0010 & -0.0391 & 12045 \\
60212.3390 & 0.0011 & -0.0413 & 11342 & 60548.3578 & 0.0009 & -0.0452 & 12046 \\
60215.1994 & 0.0007 & -0.0447 & 11348 & 60548.8386 & 0.0015 & -0.0417 & 12047 \\
60215.6874 & 0.0007 & -0.0341 & 11349 & 60549.3169 & 0.0010 & -0.0406 & 12048 \\
60216.1592 & 0.0006 & -0.0395 & 11350 & 60549.7939 & 0.0011 & -0.0410 & 12049 \\
60216.6396 & 0.0009 & -0.0364 & 11351 & 60550.2738 & 0.0010 & -0.0384 & 12050 \\
60217.1129 & 0.0006 & -0.0404 & 11352 & 60550.7525 & 0.0005 & -0.0370 & 12051 \\
60222.8458 & 0.0007 & -0.0352 & 11364 & 60551.7067 & 0.0005 & -0.0374 & 12053 \\
60223.3253 & 0.0005 & -0.0329 & 11365 & 60555.9996 & 0.0004 & -0.0403 & 12062 \\
60223.7956 & 0.0009 & -0.0400 & 11366 & 60556.4755 & 0.0018 & -0.0417 & 12063 \\
60225.7064 & 0.0019 & -0.0384 & 11370 & 60557.4252 & 0.0004 & -0.0465 & 12065 \\
60226.1872 & 0.0018 & -0.0348 & 11371 & 60557.9112 & 0.0013 & -0.0379 & 12066 \\
60226.6657 & 0.0005 & -0.0337 & 11372 & 60558.3864 & 0.0011 & -0.0400 & 12067 \\
60227.1423 & 0.0008 & -0.0344 & 11373 & 60558.8624 & 0.0019 & -0.0413 & 12068 \\
    \hline
    \end{tabular}
    \begin{flushleft}
        $^\dagger$: VarAstro data. All the other data are from TESS.
    \end{flushleft}
    \label{tab:o-c2}
\end{table}

\end{document}